\documentclass[a4paper,11pt]{article}
\usepackage{a4wide}
\usepackage{amsmath,amssymb,amsthm,mathtools,array,booktabs}
\usepackage{graphicx,hyperref}
\usepackage{xcolor}
\usepackage{cancel}
\usepackage{tikz}
\usepackage{tikz-cd}
\usepackage{amscd}
\usepackage{latexsym,cite}
\usepackage{subcaption}
\usepackage{ytableau}
\usepackage{scalerel}
\usepackage{calc}
\usepackage{authblk}
\input xy
\xyoption{all}

\hypersetup{
  unicode,
  bookmarksnumbered,
  linktoc = all,
  pdfborderstyle = {/S/U/W 0.5}
}

\let\Re\relax

\DeclareMathOperator{\Re}{Re}

\DeclarePairedDelimiter{\abs}{\lvert}{\rvert}
\DeclareMathAlphabet\mathbfcal{OMS}{cmsy}{b}{n}

\numberwithin{equation}{section}

\title{\textbf{B-brane transport in nonabelian GLSMs for $K_{Gr(2,N)}$}}
\author[2,1]{Jirui Guo\footnote{jrkwok@tongji.edu.cn}}
\author[1,3,4]{Mauricio Romo\footnote{mromoj@simis.cn}}
\author[1]{Lucy Smith\footnote{lucy.cr.smith@gmail.com}}
\affil[1]{Yau Mathematical Sciences Center, Tsinghua University, Beijing 100084, China}
\affil[2]{School of Mathematical Sciences and Institute for Advanced Study,\protect\\ Tongji University, Shanghai 200092, China}
\affil[3]{Center for Mathematics and Interdisciplinary Sciences, Fudan University, Shanghai, 200433, China}
\affil[4]{Shanghai Institute for Mathematics and Interdisciplinary Sciences (SIMIS), Shanghai, 200433, China}
\date{\today}

\begin{document}
\maketitle

\begin{abstract}
\noindent
We study the properties of B-branes in a class of nonabelian GLSMs realizing the canonical line bundle $K_{Gr(2,N)}$ in their geometric phase. By analysing the hemisphere partition function, i.e. B-brane central charge, we propose a grade restriction rule and the corresponding window categories for a specific class of paths between phases. We find very striking differences between the cases of $N$ even and $N$ odd. In particular, for the case of $N$ even, we suggest that more than one window category can be possible, for a fixed path. A detailed computation of the open Witten index and some monodromies provides evidence for our proposal for window categories. In addition, we make some remarks about B-branes on the strongly coupled phase, for the case $N=4$, based on our window proposal.
\end{abstract}
\newpage
\tableofcontents
\setcounter{footnote}{0}

\section{Introduction}

Supersymmetric boundary conditions in $\mathcal{N}=(2,2)$ gauged linear sigma model (GLSM) has been an active arena for new discoveries in mathematics and physics, with several questions still remaining open. The most well-understood class of boundary conditions correspond to the so called B-type or B-branes. They are defined as preserving a certain  subalgebra of the $\mathcal{N}=(2,2)$ supersymmetry \cite{Hori:2000ic}. In the present work we will restrict to nonanomalous GLSMs, i.e. models whose axial and vector R-symmetries are not anomalous. Such models enjoy several properties. The most relevant for us are: the RG flow to $\mathcal{N}=(2,2)$ superconformal field theories (SCFT) and a subspace $\mathcal{M}_{K}$, usually referred as the (quantum) K\"ahler moduli space of a GLSM, of their space of coupling constants can be identified with marginal deformations of the SCFT by elements of the $(a,c)$ ring \cite{Lerche:1989uy} (see \cite{Knapp:2020oba} for a review). Because of this, small deformations of the GLSM around points in $\mathcal{M}_{K}$ leave the space of B-branes unaffected, i.e. the B-branes are insensitive to small changes of the parameters $\mathcal{M}_{K}$.

It is well known that a real slice of $\mathcal{M}_{K}$ is divided into chambers called phases \cite{Witten_1993} and in each phase IR fixed points may require a different description. One important question then is how the space of B-branes changes as we move around different phases. This problem was solved for abelian GLSMs in \cite{Herbst:2008jq} and a proposal is presented in \cite{Hori:2013ika} on how one can handle nonabelian GLSMs (with further progress in \cite{EHKR,Eager2017}). In this paper we will focus on a family of nonabelian models with the following characteristics:
\begin{itemize}
 \item $\mathrm{dim}_{\mathbb{C}}\mathcal{M}_{K}=1$, the gauge group is $G=U(2)$ and there exists a phase where the IR fixed point can be described by a nonlinear sigma model (NLSM) with target space $K_{Gr(2,N)}$ (i.e. a geometric phase).

 \item Besides the geometric phase, we have one more phase which is strongly coupled (in the sense of \cite{Hori:2006dk}). Moreover, this phase, in the case of $N$ even, is irregular \cite{Hori:2011pd}, in the sense that there exists a, in principle, noncompact Coulomb branch.
\end{itemize}

The B-branes in GLSMs are matrix factorizations of the superpotential, which constitute an $A_\infty$ category and are important tools in analyzing the dynamics of GLSMs \cite{Herbst:2008jq, Hori:2013ika}. These B-branes are characterized by algebraic and symplectic data (as reviewed in Sec. \ref{sec:section2}). Their algebraic data includes a representation $\rho_{M}:G\rightarrow GL(M)$ of the gauge group $G$. The way a B-brane changes as we move between phases is termed B-brane transport and only certain B-branes can move across phases: the ones whose $\rho_{M}$ weights satisfy certain numerical constraints. These constraints are different when we cross along different paths and they are known as the grade restriction rule \cite{Herbst:2008jq,ballard2019variation,halpern2015derived,segal2011equivalences}. Determining these constraints for each phase boundary is equivalent to defining a subcategory $\omega$ of the category of B-branes $MF_{G}(W)$ in the GLSM (see Sec. \ref{sec:section2}) known as a `window category'. In the present work, we propose constraints, i.e. window categories for paths connecting the geometric phase with the strongly-coupled phase of our model and we provide several consistency checks for them. We call them windows for straight paths\footnote{Since there exists several other possible paths, for which we do not compute the grade restriction rule.} (see Sec. \ref{sec:GRR}). The window categories $\omega$ for the $K_{Gr(2,N)}$ GLSM is given by a collection of the UV lifts of vector bundles over $K_{Gr(2,N)}$. These are always given by pullbacks of bundles over $Gr(2,N)$ under the projection $\pi: K_{Gr(2,N)} \rightarrow Gr(2,N)$. Indeed we only need to consider the bundles $\pi^{*}S^{(l)}:=\pi^{*}\mathrm{Sym}^{l}S$, where $S$ is the rank $2$ tautological bundle over $Gr(2,N)$ and their twists by $\pi^*\mathrm{det}S$. 
Let $\mathcal{W}_{\mu}$ be the GLSM brane whose Chan-Paton space is the irreducible $U(2)$-module with highest weight $\mu$, then the IR image of $\mathcal{W}_{\mu}$ in the geometric phase is $\pi^*L_\mu S$, where $L_\mu$ is the Schur functor corresponding to the weight $\mu$.
A summary of our results, regarding the window categories, is then the following
(the coordinate in $\mathcal{M}_{K}$ is denoted $t=\xi-i\theta$, as described in Sec. \ref{sec:section2}):
\begin{itemize}
 \item For $N=4$, one can choose between two possible window categories i.e. two different constraints to go between phases $\xi\gg 1$ and $\xi\ll-1$. These are given, for \\
$\theta\in(2\pi l,2\pi(l+1)), l\in\mathbb{Z}$, by:
 \begin{eqnarray}
\omega^{(1)}_{2l}&:=&\{\mathcal{W}_{(-2,-2)},\mathcal{W}_{(-1,-1)},\mathcal{W}_{(0,0)},\mathcal{W}_{(1,1)},\mathcal{W}_{(0,-1)},\mathcal{W}_{(-1,-2)}\}\otimes \mathcal{W}_{-l,-l}, \nonumber\\
\omega^{(2)}_{2l}&:=&\{\mathcal{W}_{(-2,-2)},\mathcal{W}_{(-1,-1)},\mathcal{W}_{(0,0)},\mathcal{W}_{(1,1)},\mathcal{W}_{(1,0)},\mathcal{W}_{(0,-1)}\}\otimes \mathcal{W}_{-l,-l}.
\end{eqnarray}
We find that either option is equally valid. However in Secs. \ref{G24Mon} and \ref{sec:peculiarity} we show that, in order to compute the monodromy, i.e. the transport of branes around singular points in $\mathcal{M}_{K}$, we have to choose either $\omega^{(1)}$ or $\omega^{(2)}$ depending on the direction of the loop.

\item For $N=2n$, $n\in\mathbb{Z}_{>2}$, we have the option of choosing $2^{n}$ windows. In Sec. \ref{sec:evenresol}, we propose that all these choices correspond to subwindows of a larger one that we denote $\hat{\omega}_{l}$. Each subwindow is obtained from $\hat{\omega}_{l}$ by quotienting by relations imposed by certain empty branes (in the $\xi\gg 1$ phase). Here we describe one possible choice, denoted by $\omega'_{l}$, for $\theta\in(l\pi,\pi(l+1))$. We can write it as follows. Define $s:=\lfloor\frac{l+1}{2}\rfloor$. Then the corresponding window $\omega'_{l}$ is given by:
\begin{enumerate}
 \item For $l$ even
 \begin{eqnarray}
\omega'_{l}&=&\left\{\mathcal{W}_{(r+j-\lfloor\frac{r}{2}\rfloor,~j-\lfloor\frac{r}{2}\rfloor)}:-n\leq j\leq n-1,0\leq r\leq n-2\right\}\otimes\mathcal{W}_{(-s,-s)}\nonumber \\
&\cup&\left\{\mathcal{W}_{(n-1+j-\lfloor\frac{n-1}{2}\rfloor,~j-\lfloor\frac{n-1}{2}\rfloor)}:-n\leq j\leq -1 \right\}\otimes\mathcal{W}_{(-s,-s)}.
\end{eqnarray}
\item For $l$ odd, we define $\delta_{r}:=0$ if $r$ is even and $\delta_{r}:=1$ if $r$ is odd. Then
\begin{eqnarray}
\omega'_{l}&=&\left\{\mathcal{W}_{(r+j-\lfloor\frac{r}{2}\rfloor-\delta_{r},~j-\lfloor\frac{r}{2}\rfloor-\delta_{r})}:-n+1\leq j\leq n,0\leq r\leq n-2\right\}\otimes \mathcal{W}_{(-s,-s)}
\nonumber\\
&\cup&\left\{\mathcal{W}_{(n-1+j-\lfloor\frac{n-1}{2}\rfloor-\delta_{n-1},~j-\lfloor\frac{n-1}{2}\rfloor-\delta_{n-1})}:-n+1\leq j\leq 0\right\}\otimes \mathcal{W}_{(-s,-s)}.
\end{eqnarray}
\end{enumerate}
In this case, opposite to the $N=4$ case, any choice between the $2^{n}$ are on equal footing. We can compute the monodromies encircling $\theta=0$ mod $2\pi$
 or $\theta=\pi$ mod $2\pi$ just by choosing the initial window, regardless of the direction. The choice is argued, in Sec. \ref{EvenNBraneGen}, to be independent of the window we choose.


\item For $N=2n+1$, $n \in \mathbb{Z}_{>0}$, we have a unique window. We can write them as follows. When $\theta\in(l\pi,\pi(l+1))$ define $s:=\lfloor\frac{l+1}{2}\rfloor$. Then the corresponding window $\omega_{l}$ is given by:
\begin{enumerate}
 \item For $l$ even, we define $\delta_{r}:=0$ if $r$ is even and $\delta_{r}:=1$ if $r$ is odd. Then
 \begin{equation}
\omega_{l}= \left\{\mathcal{W}_{(r+j-\lfloor\frac{r}{2}\rfloor-\delta_{r},~j-\lfloor\frac{r}{2}\rfloor-\delta_{r})}:-n\leq j\leq n,0\leq r\leq n-1\right\}\otimes \mathcal{W}_{(-s,-s)}.
\end{equation}
\item For $l$ odd
 \begin{equation}
\omega_{l}= \left\{\mathcal{W}_{(r+j-\lfloor\frac{r}{2}\rfloor,~j-\lfloor\frac{r}{2}\rfloor)}:-n\leq j\leq n,0\leq r\leq n-1\right\}\otimes \mathcal{W}_{(-s,-s)}.
\end{equation}
\end{enumerate}
In Sec. \ref{OddNBraneGen} we argue about the general structure of the monodromies in this case.
\end{itemize}

We provide evidence for the windows described above by computing the open Witten index $\chi(\mathcal{B}_{1},\mathcal{B}_{2})$ between two grade-restricted B-branes in both $\xi\gg 1$ and $\xi\ll -1$ phases and show that both computations agree (see Sec. \ref{sec:open}). In addition, for the cases $N=4,5$, we can identify the monodromy implemented by straight paths with spherical twists. We also check the consistency of this fact with the perturbative part of the hemisphere partition function in Sec \ref{sec:monodromy}.

The $\xi\ll -1$ phase, being strongly coupled, possesses a challenge to analyze its low-energy dynamics. Based on the observations in \cite{Lerche:1989uy} and a classical analysis of the vacuum equations, we conjecture, as suggested to us by \cite{EH}, that this phase should reduce to a $\mathbb{Z}_{N}$ orbifold of some noncommutative resolution of the affine cone over the Grassmannian: $CGr(2,N)$. The analysis of the $\xi\ll -1$ phase is particularly complicated in the case of $N$ even because of the possibility of a noncompact Coulomb branch (see \ref{sec:section2} for more details). Nevertheless we propose in Sec. \ref{sec:negphase} some semiorthogonal decomposition for the category of B-branes on the $\xi\ll -1$ phase for the case $N=4$ based on the expectation that it should be equivalent to $D^{b}Coh(K_{Gr(2,4)})$, being the latter, the category of B-branes on the $\xi\gg 1$ phase and our proposal for windows.

This paper is organized as follows. In Sec. \ref{sec:section2}, we introduce some general aspects of the $K_{Gr(k,N)}$ GLSM including the location of the singular points. In Sec. \ref{sec:Btransport} we review the definition of B-branes in GLSMs, determine the grade restriction rule for straight paths, and the window categories. We explain why, in the even $N$ case, there is an ambiguity when choosing the windows and we propose a solution to such a puzzle. In Sec. \ref{sec:open} we propose a formula for the computation of the open Witten indices\footnote{Our formula is based on the localization result of \cite{Hori:2013ika}. The only missing ingredient is the integration contour, which is the only new ingredient in our proposal (the contour is left ambiguous in \cite{Hori:2013ika} and only presented explicitly in examples).} at different phases of the $K_{Gr(2,N)}$ GLSMs and present various examples of its consistency with the grade restriction rule as evidence for it. In Sec. \ref{sec:monodromy} we compute the monodromy using straight paths in various examples. We also present some general structure of such monodromies. Finally, in Sec. \ref{sec:negphase}, we investigate the IR behaviour of the $\xi\ll -1$ phase from the point of view of B-branes, for the $N=4$ case.

\section{\label{sec:section2}GLSM for $K_{Gr(k,N)}$}

In this section, we present the basic ingredients of the GLSM for $K_{Gr(k,N)}$, including the phases, the B-branes and the singularities on the moduli space $\mathcal{M}_{K}$.

\subsection{The model}\label{sec:setup}

The GLSM for $K_{Gr(k,N)}$ has $U(k)$ gauge group, $N$ chiral fields $\Phi_i$,
$i=1,\cdots,N$, in the fundamental representation  $\square:=\mathbf{k}$ and
one chiral field $P$ in the $\det^{-N}$ representation, and vanishing
superpotential. The matter content can be summarised as follows:
\begin{equation}\label{mattercontent}
\centering
\begin{tabular}{c|ccccc}
        & $\Phi_1$ & $\Phi_2$ & $\cdots$ & $\Phi_N$ & $P$  \\ \hline
 $U(k)$ & $\square$ & $\square$ & $\cdots$ &  $\square$  & $\det^{-N}$
\end{tabular}
\end{equation}

Let $\phi_i^\alpha$ and $p$ be the scalar component fields of $\Phi_i^\alpha$ and $P$ respectively ($\alpha=1,2,\cdots,k$ is the color index), then the D-term equations read
\begin{align}
	D^\alpha_\beta = \sum_i \phi^\alpha_i \bar{\phi}_{i \beta} - \delta^\alpha_\beta (N|p|^2 + \xi) = 0,
\end{align}
where $\xi\in \mathbb{R}$ denotes the FI-parameter associated with the
determinant $U(1)$ subgroup of $U(k)$.
The diagonal terms $D^\alpha_\alpha$ set the norm of the $\phi^\alpha$ vectors,
\begin{align}\label{norm}
	\sum_{i=1}^N \phi^\alpha_i \bar{\phi}_\alpha^i = N|p|^2 + \xi, \quad \forall \alpha,
\end{align}
while the off-diagonal terms require them to all be orthogonal,
\begin{align}\label{ortho}
	\sum_{i=1}^N \phi^\alpha_i \bar{\phi}_\beta^i = 0, \quad \forall \alpha \ne \beta.
\end{align}

The D-terms split the moduli space into two phases: a positive phase with $\xi \gg 1$ and a negative phase\footnote{Throughout this work we write $\xi \gg 1$ and $\xi \ll -1$ to emphasize that we are choosing $|\xi|$ very large, and away from the singularities coming from the Coulomb branch.} with $\xi \ll -1$. There is a classical phase boundary located at $\xi=0$.

In the positive phase, $\xi \gg 1$, then Eq. \eqref{norm} and \eqref{ortho} tell
us that $\{ \phi^\alpha ~|~ \alpha=1,2,\cdots,k \}$ is an orthonormal basis of a
$k$-dimensional subspace inside $\mathbb{C}^N$. Upon taking the quotient by
$U(k)$, $\phi^\alpha_i$ becomes the Stiefel coordinates of the Grassmannian
$Gr(k,N)$, the set of $k$-dimensional subspaces in $\mathbb{C}^N$. Because $p$
transforms in the $\det^{-N}$ representation, it becomes the coordinate along
the fibre of the line bundle $(\det S^\vee)^{\otimes N}$, which is exactly the
canonical line bundle, where $S$ is the tautological bundle over $Gr(k,N)$.
Therefore, the target space of the low-energy NLSM in this phase becomes
\[
X = K_{Gr(k,N)} = \mathrm{Tot}\left( (\det S^\vee)^{\otimes N} \rightarrow Gr(k,N) \right),
\]
which is a non-compact Calabi-Yau manifold.

In the negative phase, $\xi \ll -1$, $p$ can never be zero lest the norms of the $\phi^\alpha$ $N$-vectors become negative. However, they are, in this phase, allowed zero norms and thus the vacuum expectation values $\phi^\alpha_i =0$, $\forall \alpha,i$, are available. In this case, the vev of $p$ becomes fixed to a point, specifically
\begin{align}
	\langle p \rangle = \sqrt{\frac{-\xi}{N}},
\end{align}
which breaks the gauge group to a subgroup $H \subset U(k)$ whose matrices $M$
all have the feature $(\det M)^N=1$. The resulting effective theory is thus a
gauge theory with gauge group $H$, massless chiral fields $\Phi^\alpha_i$ and
vanishing superpotential. As analyzed in \cite{Lerche:2001vj}, the classical solution to the D-term equations, upon quotient by the gauge group, is given by a
$\mathbb{Z}_N$-orbifold of the affine cone of $Gr(k,N)$ under the Pl\"ucker
embedding into $\mathbb{P}^{{N \choose k}-1}$. This space, being singular, cannot be the whole story. We analyze this phase, for the case $N=4$ in Sec. \ref{sec:negphase}, from the point of view of B-branes.

In the case at hand, the B-brane boundary conditions of the GLSM are simply
Chan-Paton spaces carrying representations of $U(k)$, as we will review in
more detail in Sec. \ref{sec:Btransport}. Let $\mathcal{W}_\mu$ be such a
brane with the irreducible representation corresponding to the highest weight $\mu =
(\mu_1,\mu_2,\cdots,\mu_k)$, where $\mu_i \in \mathbb{Z}$ satisfying $\mu_1 \geq
\mu_2 \geq \cdots \geq \mu_k$ (see for example \cite{Taylor}). Then, for $\xi \gg
1$, $\mathcal{W}_\mu$ becomes the sheaf $\pi^*(L_\mu S)$ under the RG flow,
where $\pi$ is the projection $\pi: K_{Gr(k,N)} \rightarrow Gr(k,N)$ and
$L_\mu$ is the Schur functor corresponding to $\mu$\footnote{For a detailed
introduction of Schur functors, see Chapter 2 of \cite{weyman_2003}.}. For
example, in the case of $k=2$, the IR image of $\mathcal{W}_{(2,0)}$ is
$\pi^*(\mathrm{Sym}^2 S)$ and the IR image of $\mathcal{W}_{(1,1)}$ is
$\pi^*(\det S)$. 

\subsection{Singular points}

On the complexified K\"ahler moduli space $\mathcal{M}_K$ parametrized by the
FI-theta parameter $t = \xi- i \theta$, where $\theta$ is the theta-angle, the
phase boundary receives quantum correction and becomes singular points. Such
points are removed from the moduli and so they become punctures in
$\mathcal{M}_K$. At singular points, non-compact directions along the
Coulomb branch open up. In the nonabelian theories, there are typically more
than one singular points as pointed out in \cite{Hori:2006dk}.

To find the locations of the singular points, let us study the vacuum structure
on the Coulomb branch. If, for a specific value of $t$, the vacuum equations on
the Coulomb branch have a space of solutions with positive dimension, then that
value of $t$ corresponds to a singular point.

The twisted superpotential on the Coulomb branch reads\footnote{We remark that
the terms $\pi i \sum_{a < b} (\sigma_a -
\sigma_b)$ come from the W-bosons and was missing in \cite{Hori:2006dk} but
corrected in \cite{Hori:2011pd} and re-derived by a localization computation in \cite{Hori:2013ika}.} \cite{hori2019notes}
\begin{equation}\label{superpotential}
\begin{split}
\widetilde{W} = &-t \sum_{a=1}^k \sigma_a + \pi i \sum_{a < b} (\sigma_a -
\sigma_b) -N \sum_{a=1}^k \sigma_a \left[ \log\left( \sigma_a/\Lambda \right) -
1 \right] \\ &+ N \left( \sum_{a=1}^k \sigma_a \right) \left[ \log\left( -N
\sum_{a=1}^k \sigma_a/\Lambda \right) - 1 \right],
\end{split}
\end{equation}
where $\sigma_{a}$ are coordinates on the (complexified) maximal torus of
$U(k)$ and $\Lambda$ denotes a UV cut-off\footnote{We use natural units, in which $\sigma$ has unit of energy.}. In deriving (\ref{superpotential}), it is assumed that $|\sigma_{a}|\gg 1$
for all $a$ and that $U(k)$ is completely broken to its maximal torus, hence
$\sigma_{a}\neq \sigma_{b}$ for $a\neq b$. As mentioned above, the singular
points can be determined by requiring that the vacuum equations, which are given
by
\[
\exp\left( \frac{\partial \widetilde{W}}{\partial \sigma_a} \right)=1,\quad
a=1,\cdots, k,
\]
have a space of solutions with positive dimension.
From \eqref{superpotential},
\[
\frac{\partial \widetilde{W}}{\partial \sigma_a} = -t + \pi i (k-2a+1) - N \log
\sigma_a + N \log(-N \sigma),\qquad \sigma:=\sum_{a=1}^{k}\sigma_{a}
\]
therefore the equations of vacuum read
\begin{equation}\label{EOM}
(-1)^{k+1} \sigma_a^N = (-N \sigma)^N \exp(-t),\quad a=1,\cdots,k,
\end{equation}
which tells us that $(\sigma_a/\sigma_b)^N = 1$ for all $a$ and $b$.

Now we specialize to $k=2$. In this case, the identity $\sigma_1^N = \sigma_2^N$ induces $\sigma_2 = \omega \sigma_1$, where $\omega^N = 1$. Because of the condition on the excluded loci $\sigma_1 \neq \sigma_2$, we have $\omega \neq 1$. Moreover, $\omega \neq -1$, otherwise Eq. \eqref{EOM} would lead to $\sigma_1 = \sigma_2 = 0$. Another way to see that $\omega \neq -1$ is given in \cite{Hori:2006dk}: if $\sigma_{1}+\sigma_{2}=0$ this choice is not generic, since it will keep the field $P$ massless. From Eq. \eqref{EOM} and $\sigma_2 = \omega \sigma_1$, we get
\[
(-N (1+\omega) \sigma_1)^N \exp(-t) = - \sigma_1^N,
\]
i.e. the singular points are located at
\begin{equation}\label{singk2}
\exp(-t) = - (-N (1+\omega))^{-N},
\end{equation}
where $\omega^N=1$ and $\omega \neq \pm 1$.  Let's denote $\omega = \exp(2 \pi i l/N)$ with $l \in \mathbb{Z}$, then
\[
\mathrm{Arg} \left( \frac{1}{1+\omega} \right) = - \frac{\pi l}{N}.
\]
Then Eq. \eqref{singk2} implies that the singular points have theta-angle
\begin{equation}\label{singtheta}
\theta \equiv N \mathrm{Arg} \left( \frac{1}{1+\omega} \right) + (N+1) \pi \equiv (N+1-l) \pi \quad \mathrm{mod}~ 2 \pi ,
\end{equation}
where, due to the constraints on $\omega$, the integer $l$ satisfies $l \neq 0$, and $ -\frac{N-1}{2} < l \leq \frac{N-1}{2}$ for odd $N$, $ -\frac{N}{2}+1 < l \leq \frac{N}{2}- 1$ for even $N$.

The locations on the theta-angle line for the first few values of $N$ are given in Table \ref{SingPointTable}. Repeated entries indicate that multiple singular points are found at the same value of $\theta$, but sit at different locations in the $r$-direction.

\begin{table}
	\begin{center}
		\begin{tabular}{cc}
			\hline
			$N$ & $\theta$ \\
			\hline
			4 & 0 \\
			5 & 0, $\pi$ \\
			6 & 0, $\pi$ \\
			7 & 0, 0, $\pi$ \\
			8 & 0, $\pi,\pi$ \\
			\hline	
		\end{tabular}
		\caption{The location of singular points on the moduli space of various models of $K_{Gr(2,N)}$.}
		\label{SingPointTable}
	\end{center}
\end{table}

In the following we will study how B-type boundary conditions change as we
deform $t$ along a path in $\mathcal{M}_{K}$, a process termed B-brane
transport and whose relevant aspects will be reviewed in Sec.
\ref{sec:Btransport}. In this current work, we will be concerned with B-brane
transport along straight paths given by $r\in \mathbb{R}$ and constant
$\theta$ with $\theta\in (0,\pi)+2\pi \mathbb{Z}$ or
$\theta\in
(\pi,2\pi)+2\pi \mathbb{Z}$. When considering boundary theories, the
correlators are no longer periodic functions of $\theta$, hence we have to work
on a covering of $\mathcal{M}_{K}$, and consider $\theta\in\mathbb{R}$.
Therefore we have two infinite families of paths we can consider, whenever
$N>4$ (for $N=4$ there is a single solution to \eqref{EOM}). The boundaries of
the intervals where the paths are located are chosen to be $\theta=0,\pi$,
because the singularities are located there. It is also natural to consider more
complicated paths, for $N>6$, that are curved, and goes in between a pair of
singularities, crossing from $(0,\pi)+2\pi \mathbb{Z}$ to $(\pi,2\pi)+2\pi
\mathbb{Z}$. B-brane transport along these curved paths is much more
challenging to analyze \cite{EHKR} and we will not consider them in the present
work.

\subsection{Regularity of the $\xi\ll -1$ phase}\label{sec:regularity}

In this subsection we comment on the phenomenon of irregular phases in GLSMs, as it applies to the present $K_{Gr(k,N)}$ model. Irregular phases can appear in nonabelian GLSMs, and correspond to phases where the existence of a noncompact Coulomb branch is allowed by the twisted potential equations \cite{Hori:2011pd}. Further examples, including geometric models with irregular phases appear in \cite{Hori:2013gga}.

In $K_{Gr(k,N)}$ models, the $\xi\ll-1$ phase has an unbroken gauge group $H$ with an $SU(k)$ subgroup\footnote{More precisely the unbroken gauge group in the $\xi\ll-1$ is the subgroup of elements $g\in U(k)$ satisfying $(\det g)^{N}=1$.}. Then the twisted superpotential of such a phase can be computed, assuming $|\sigma_{a}|\gg 1$ for all $a$ and in the branch where it is completely broken. This gives precisely (\ref{superpotential}) but with the restriction $\sum_{a=1}^{k}\sigma_{a}=0$. Explicitly,
\begin{equation}\label{twisted_neg}
\widetilde{W}_{-} := 2\pi i\sum_{a=1}^{k-1}(k-a)\sigma_{a} -N \sum_{a=1}^{k-1} \sigma_a \left[ \log\left( \sigma_a/\Lambda \right) -
1 \right] - N \left( \sum_{a=1}^{k-1} \sigma_a \right) \left[ \log\left( -
\sum_{a=1}^{k-1} \sigma_a/\Lambda \right) - 1 \right].
\end{equation}
Then, the vacuum equations read:
\begin{equation}\label{veq_twisted_neg}
\sigma_{a}^{N}-(-1)^{N}\left(\sum_{a=1}^{k-1} \sigma_a\right)^{N}=0,\qquad a=1,\ldots,k-1,
\end{equation}
whenever $k|N$ we can find a solution to (\ref{veq_twisted_neg}), given by
\begin{equation}
\sigma_{a}=\omega_{k}^{a}\sigma,\qquad \omega_{k}=e^{\frac{2\pi i}{k}},
\end{equation}
where $\sigma$ is arbitrary. This solution is in agreement with the genericity conditions used to derive (\ref{twisted_neg}). Hence, we conclude that the phase $\xi\ll -1$ of the $K_{Gr(k,N)}$ model is irregular\footnote{The same conclusion was reached in \cite{Hori:2006dk}. The inclusion of W-bosons does not affect the upshot.} whenever $k|N$. In the particular case of $k=2$ and $N$ odd, it can be checked that no solution to (\ref{veq_twisted_neg}) exists, therefore these models are regular.

\section{\label{sec:Btransport}B-brane transport and window categories}

In this section, we study the grade restriction rule, window category and brane
transport for the $U(2)$ theory engineering $K_{Gr(2,N)}$. We start with a short
review on the general features of brane transport.

\subsection{A short review of B-branes on GLSMs}

For a GLSM with gauge group $G$, matter fields in the representation\footnote{In this paper we will only be concerned with nonanomalous GLSMs.} $\rho_m: G
\rightarrow \mathrm{SL}(\mathcal{V})$ and superpotential $W$,
a B-brane is described by a pair of objects $(\mathcal{B},L_{t})$ with $\mathcal{B}=(M,\rho_{M},R_{M},\mathbf{T})$, where:
\begin{itemize}
\item $M$ is the Chan-Paton vector space, which is a $\mathbb{Z}_{2}$-graded,
finite dimensional free $\mathrm{Sym}(\mathcal{V}^\vee)$ module.
\item $\rho_M$ and $R_M$ are even representations of the gauge and the (vector) R-charge representations respectively.
\item $\mathbf{T}$ is a matrix factorization of $W$, a $\mathbb{Z}_2$-odd endomorphism $\mathbf{T} \in \mathrm{End}^{1}_{\mathrm{Sym}(V^{\vee})}(M)$ satisfying $\mathbf{T}^{2}=W\cdot\mathrm{id}_{M}$.
\end{itemize}
They must satisfy the compatibility conditions: for all $\lambda\in U(1)_{V}$ (the vector R-symmetry) and $g\in G$,
  \[
    \begin{aligned}
      &R_{M}(\lambda)\mathbf{T}(R(\lambda)\phi)R_{M}(\lambda)^{-1} = \lambda \mathbf{T}(\phi) , \\
      &\rho_{M}(g)^{-1}\mathbf{T}(\rho_{m}(g)\cdot \phi)\rho_{M}(g) = \mathbf{T}(\phi) .
    \end{aligned}
  \]
$L_{t}$ is a profile for the vector multiplet scalar, it consists of a
gauge-invariant middle-dimensional subvariety of the complexified Lie algebra of
$G$. Details on how $L_{t}$ is defined can be found in \cite{Hori:2013ika} and
more recent developments in \cite{EHKR,Eager2017}. The objects $(\mathcal{B},L_{t})$ form
a category which we denote
\[
    \begin{aligned}
     MF_{G}(W).
    \end{aligned}
  \]
This category represents the B-branes on a GLSM and was defined more precisely
in \cite{Herbst:2008jq,Hori:2013ika}, however, when we are working on a
particular phase, the data $L_{t}$ can be ignored. More precisely, it is
expected that we can fix an $L_{t}$ that suits every
$\mathcal{B}$. A mathematical description of this category is given in
\cite{ballard2019variation}.

Brane transport refers to the image of a GLSM B-brane after we deform the
parameters $t$ along a smooth path on the K\"ahler moduli space from one phase
to another. In the Calabi-Yau (CY) case
(also known as the nonanomalous case, meaning that the axial U(1) R-symmetry is
anomaly free) in which the FI parameter is marginal, there are a number of
singular points on the K\"ahler moduli space (parameterized by the FI parameters)
between the two phases, so the path along which the brane is transported must
avoid these singularities. Given such a path, only the branes satisfying the
grade restriction rule can be smoothly transported from one phase to the other.
The branes satisfying the grade restriction rule constitute a subcategory of
the category of matrix factorizations of the GLSM, which is called the window
category \cite{Herbst:2008jq}.

The grade restriction rule amounts to the convergence of the hemisphere partition function along the path of transport. The hemisphere partition function can be computed by a localization formula on the Coulomb branch \cite{Hori:2013ika}:
\begin{equation}\label{Disk}
	Z_{D^2}(\mathcal{B}) = C \int_{L_{t}} d^{l_G} \sigma \;
e^{it(\sigma)}f_{\mathcal{B}}(\sigma) \prod_{\alpha>0} \alpha(\sigma)
\text{Sinh}(\pi \alpha(\sigma)) \prod_{j=1}^{\mathrm{dim}\mathcal{V}}
\Gamma(iQ_j(\sigma) + \frac{1}{2}R_j),
\end{equation}
where $C$ is a constant factor, the first product in the integrand is over all the positive roots, the second product is over all the matter fields with weight $Q_j$ and R-charge $R_j$. The brane factor is
\begin{align}
	f_{\mathcal{B}}(\sigma) =\text{Tr}_M(R_{M}(e^{i \pi})\rho_{M}(e^{2\pi
\sigma})),
\end{align}
the contour of integration coincides with $L_t$ on the  complexified
Lie algebra of $G$ and $l_{G}$ is the rank of $G$.

For a brane to be successfully transported from one phase to another, this
integral must be convergent for the entire journey across the moduli space with
some choice of the contour $L_{t}$. This places restrictions on the charges
that a brane may carry as the contour is deformed during transport across the
borders of the moduli space. Therefore the grade restriction rule can be
determined by studying the asymptotic behaviour of the integrand of
\eqref{Disk}.


\subsubsection{B-branes on $K_{Gr(k,N)}$ GLSMs}\label{sec:BbranesonGr}

The B-branes for the GLSMs presented in Sec. \ref{sec:section2} take a very
simple form. Since these models have vanishing superpotential $W\equiv 0$, the
B-branes are characterized by elements from the derived category of
($\mathbb{Z}_{2}$-graded) $U(k)$-modules. Explicitly, any B-brane
$\mathcal{B}$ is characterized by a tuple $(M,\rho_{M},R_{M},\mathbf{T})$,
where $\mathbf{T}^{2}=0$. However, note that we still impose the condition of
$\mathbf{T}$ to have weight one under $R_{M}$-action. 

A B-brane that flows to the object
$\pi^{*}L_{\mu} S_{Gr(k,N)}$ in the $\xi\gg 1$ phase, where $\pi:K_{Gr(k,N)}\rightarrow Gr(k,N)$ is the projection and $\mu$ is a dominant integral weight of $U(k)$, is simply given by the GLSM B-brane $\mathcal{W}_{\mu}$ defined in Sec.
\ref{sec:section2}\footnote{Note that in the case at hand $MF_{G}(W)=MF_{U(k)}(0)$ so this category can be identified with the derived category of graded $U(k)$-modules $D(\mathrm{gr}-U(k))$. },
whose Chan-Paton space is the $U(k)$-module with highest weight $\mu$.
All the GLSM B-branes we will be concerned with will be given by complexes whose
factors are of the form $\mathcal{W}_{\mu}$, i.e. we will construct our B-branes
by taking successive cones over morphisms between B-branes of the type $\mathcal{W}_{\mu}$.

\subsection{The grade restriction rule for straight paths:
absolute convergence}\label{sec:GRR}

In this subsection, we compute and define the grade restriction rule for
the $U(2)$ model for $K_{Gr(2,N)}$. Our definition and computation will be
specialized to what we term `straight paths'. These are paths in
$\mathcal{M}_{K}$ that have constant $\theta$. The definition of the
grade restriction rule will be based on the absolute convergence of the hemisphere
partition function $Z_{D^2}(\mathcal{B})$. We will see in Sec.
\ref{sec:evenresol} that, for $N$ even, this seems to be unsatisfactory and we
propose a solution for this family of models. For a B-brane $\mathcal{B}$,
$Z_{D^2}(\mathcal{B})$ is given by the contour integral \eqref{Disk}. We want to
derive the condition on $(L_{t},\mathcal{B})$ for a family, parameterized by
$t$, of such contour integrals to be absolutely convergent.

Let us start with rewriting the integral \eqref{Disk} in the large $|\sigma|$
region. In that limit it will take the form
\begin{align}
	Z_{D^2}(\mathcal{B}) \sim \int_\gamma g(\sigma) \sum_m e^{F_m(\sigma)},
	\label{part1}
\end{align}
where $g(\sigma)$ is a polynomial of $\sigma$, and the sum $\sum_m$ is
identified as a sum over the weights of $\rho_{M}$. So the asymptotics are
controlled by the functions $F_m(\sigma)$. Then the condition for absolute
convergence is given by
\begin{align}
	|\sigma| \rightarrow \infty \Rightarrow
A_{m}(\sigma):=\text{Re}(F_m(\sigma)) \rightarrow -\infty, \quad \forall m.
	\label{AsymCond}
\end{align}

Writing
\begin{equation}
\sigma = \tau + i\nu,
\end{equation}
it is straightforward to compute \cite{Hori:2013ika}
\begin{equation}
\begin{split}
	A_{m}(\sigma) =& \sum_{\alpha>0} \pi |\alpha(\tau)| - \xi
(\nu) + (\theta + 2\pi q^m) (\tau) \\
	&- \sum_j \left[Q_j(\nu) \left( \text{ln}(Q_j(\sigma))-1 \right) + |Q_j(\tau)| \left( \text{tan}^{-1} \left( \frac{Q_j(\nu)}{|Q_j(\tau)|}\right) +\frac{\pi}{2} \right) \right],
\end{split}
\end{equation}
where the second line is proportional to the effective boundary energy of the
matter system\cite{Herbst:2008jq}. The first term of the first line takes the
interpretation of the boundary energy of a W-boson multiplet, while the others
comprise the classical boundary potential \cite{Hori:2013ika}.

Specializing to the models for $K_{Gr(k,N)}$, we obtain (from now on, we drop the subindex $m$ to avoid cluttering)
\begin{equation}
\begin{split}
	A(\sigma) =& \sum_{a < b} \pi \left|\tau_a-\tau_b\right| + \sum_a \left[
\theta + 2 \pi q^a -N \text{Sgn}(\tau_a)\left(
\text{tan}^{-1}\left(\frac{\nu_a}{|\tau_a|} \right) + \frac{\pi}{2} \right)
\right. \\
	&\left. - N \text{Sgn}\left(\sum_{c=1}^k \tau_c\right) \left( \text{tan}^{-1} \left( \frac{-\sum_{b=1}^k \nu_b}{\left|\sum_{c=1}^k \tau_c\right|} \right) +\frac{\pi}{2} \right) \right] \tau_a \\
	&+ \sum_{a} \left[N\left(\text{ln} \left|N \sum_{b=1}^k \sigma_b\right|-\text{ln}|\sigma_a|\right) -\xi \right] \nu_a.
\end{split}
\end{equation}

In order to determine the integration contour $\gamma$, we first identify the
poles of the integrand in \eqref{Disk}.  These poles occur at
\begin{align}
	Q_j(\sigma) = i\left(n_j + \frac{1}{2}R_j\right), \quad n_j \in
\mathbb{Z}_{\geq 0}.
\end{align}
So, in the models at hand, the poles are located at $\tau_a=0$ for any
fundamental matter contribution, or $\sum_a \tau_a = 0$ for the $P$ field
contribution. In general, the R-charges take values ${0< R_j < 2}$, so the poles
can be avoided altogether by following the `wedge conditions' that must be
imposing on a contour (also called `pole-avoiding conditions' in
\cite{Hori:2013ika}):
\begin{align}
	\tau_a =0 \quad \Rightarrow& \quad \nu_a \leq 0; \label{WedgeCond1} \\
	\sum_b \tau_b = 0 \quad \Rightarrow& \quad \sum_b \nu_b \geq 0.
	\label{WedgeCond2}
\end{align}

Specializing to $k=2$, the function $A(\sigma)$
becomes
\begin{equation}
\begin{split}
	A_{(2,N)}(\sigma) =& \pi \left|\tau_1-\tau_2\right| + \theta
\left(\tau_1+\tau_2\right) - \xi \left(\nu_1+\nu_2\right) +2 \pi q^1 \tau_1
+ 2 \pi q^2 \tau_2 \\
	&-N \left|\tau_1\right| \left( \text{tan}^{-1} \left( \frac{\nu_1}{|\tau_1|}\right) + \frac{\pi}{2}  \right) -N \left|\tau_2\right| \left( \text{tan}^{-1} \left( \frac{\nu_2}{\left|\tau_2\right|}\right) + \frac{\pi}{2}  \right) \\
	&-N\left|\tau_1+\tau_2\right| \left( \text{tan}^{-1} \left( \frac{-\nu_1-\nu_2}{\left|\tau_1+\tau_2\right|}\right) + \frac{\pi}{2} \right) \\
	&+ N\left(\nu_1+\nu_2\right) \text{ln} \left|N\left(\sigma_1+\sigma_2\right)\right|-N\nu_1 \text{ln}\left|\sigma_1\right|-N\nu_2 \text{ln}\left|\sigma_2\right|.
\end{split}
\end{equation}
We propose to use the same contour as the one found in \cite{EHKR} for the, very closely related, Pfaffian-Grassmannian model \cite{Hori:2006dk}. This contour corresponds to, for straight paths in
$\mathcal{M}_{K}$:
\begin{align}
	\nu_1 = -\nu_2 = \left(\tau_1\right)^2 - \left(\tau_2\right)^2,
	\label{Contour}
\end{align}
as this satisfies the wedge conditions of (\ref{WedgeCond1} - \ref{WedgeCond2})
automatically. Evaluating $A_{(2,N)}(\sigma)$ on
\eqref{Contour} gives
\begin{equation}
\begin{split}
	A_{(2,N)}(\sigma) =& \pi \left|\tau_1-\tau_2\right| + \theta
\left(\tau_1+\tau_2\right) +2 \pi q^1 \tau_1 + 2 \pi q^2 \tau_2 \\
	&-N \left|\tau_1\right| \left( \text{tan}^{-1} \left( \frac{\left(\tau_1\right)^2 - \left(\tau_2\right)^2}{\left|\tau_1\right|}\right) + \frac{\pi}{2}  \right) \\
	&-N \left|\tau_2\right| \left( \text{tan}^{-1} \left( \frac{\left(\tau_2\right)^2 - \left(\tau_1\right)^2}{\left|\tau_2\right|}\right) + \frac{\pi}{2}  \right) - \frac{1}{2} N \pi \left|\tau_1+\tau_2\right| \\
	&-N\left[\left(\tau_1\right)^2 - \left(\tau_2\right)^2\right] \text{ln}\left|\frac{\tau_1 + i\left[\left(\tau_1\right)^2 - \left(\tau_2\right)^2\right]}{\tau_2-i\left[\left(\tau_1\right)^2 - \left(\tau_2\right)^2\right]}\right|.
\end{split}
\end{equation}

Now we can test the asymptotic behaviour of $A_{(2,N)}$ at different regions of the $\sigma$-space. In each of the regions, the requirement $\Re(A_{(2,N)}) \rightarrow -\infty$ as $|\sigma| \rightarrow \infty$ results in a constraint on the $U(2)$ weights $(q^1,q^2)$ carried by the brane.

The analysis only needs to be performed in eight regions in $\tau$-space,
whose boundaries are defined by the four lines $\tau_{1}=\pm\tau_{2}$,
$\tau_{i}=0$, $i=1,2$ \cite{EHKR}. It is enough to analyze the asymptotics of
$A_{(2,N)}(\sigma)$ at the boundaries. As an example,  consider the case
$\tau_1=\tau_2 \equiv \tau$ and take the limit $|\tau| \rightarrow \infty$,
bringing the asymptotic function to
\begin{align}
	A_{(2,N)}(\sigma) \rightarrow& \left[\left(\frac{\theta}{\pi} + q^1 +
q^2\right)\text{Sgn}(\tau) - N\right] |\tau|.
\end{align}
The conditions \eqref{AsymCond} forces the bracketed expression to be less than
zero, which yields the following condition on the gauge charges:
\begin{align}
	-N < \frac{\theta}{\pi} + q^1 + q^2 < N.
\end{align}

Likewise, we obtain:
\begin{align}
	\tau_1=\tau_2 \equiv \tau \quad \Rightarrow& \quad \left| \frac{\theta}{\pi} + q^1 + q^2 \right| < N, \label{GRR2a} \\
	\tau_1=-\tau_2 \equiv \tau \quad \Rightarrow& \quad \left|q^1 - q^2 \right| < \frac{1}{2}(N-2), \label{GRR2b} \\
	\tau_a \equiv \tau, \tau_{a\pm 1} = 0 \quad \Rightarrow& \quad \left| \frac{\theta}{2 \pi} + q^a \right| < \frac{1}{4}(3N-2). \label{GRR2c}
\end{align}

\begin{figure}[h]
	\begin{subfigure}[b]{0.45 \linewidth}
		\includegraphics[width=\linewidth]{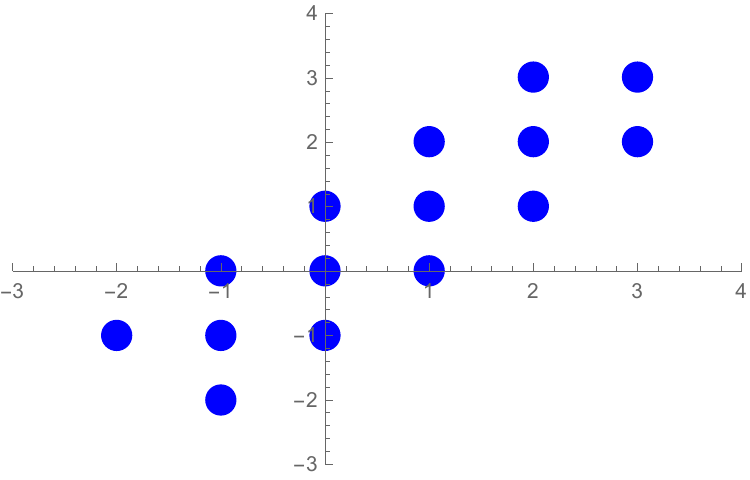}
		\caption{$\omega_{-3}$.}
      \label{CharWin25-1}
	\end{subfigure}
	\begin{subfigure}[b]{0.45 \linewidth}
		\includegraphics[width=\linewidth]{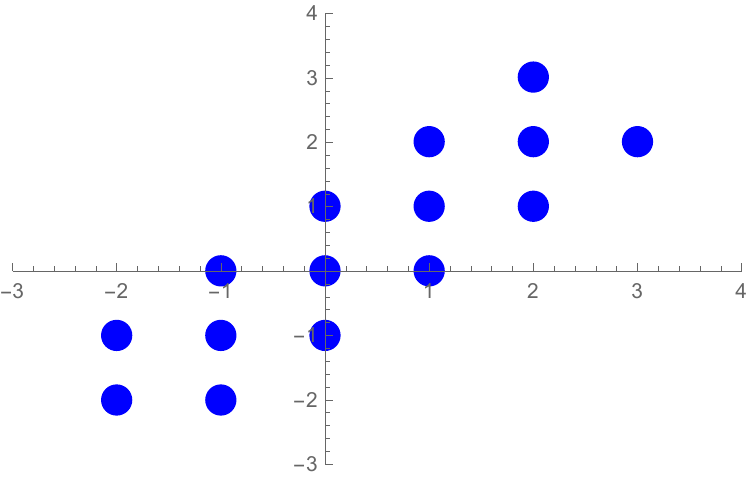}
		\caption{$\omega_{-1}$.}
      \label{CharWin25-2}
	\end{subfigure}
	\centering
	\caption{The charge restrictions for two windows of
$K_{Gr(2,5)}$. The horizontal and vertical axes label the first and
second components of the $U(2)$ weight $(q^1,q^2)$ respectively.}
	\label{G25MonWins1}
\end{figure}

As an example, Fig. \ref{G25MonWins1} illustrates two charge restrictions
for $K_{Gr(2,5)}$: ${-3\pi < \theta < -2 \pi}$ and ${-\pi < \theta < 0}$,
denoted $\omega_{-3}$ and $\omega_{-1}$ respectively. In this case, the jump at
$\theta=0,\pi$ $\mathrm{mod}$ $2\pi$ is completely consistent with the location
of the singularities in $\mathcal{M}_{K}$. We will see in Sec.
\ref{sec:evenresol} that such restrictions on the charges or weights of
$\rho_{M}$ define a category that we will term `window category' as originally
termed in \cite{Herbst:2008jq}. However, for even $N$, this category (as can be infered from the fact that, for even $N$, this theory is irregular as explained in Sec. \ref{sec:regularity}) is
incompatible with the physical expectations and we will propose a solution to
this inconsitency in Sec. \ref{sec:evenresol} \footnote{We also were informed of a work in progress finding similar results \cite{MK}.}.

\subsection{The travelling branes: choice of generators}\label{sec:trav}

With the grade restriction rule found in Sec. \ref{sec:GRR}, we can now
implement the brane transport along straight paths. For this purpose, first we
choose a set of generators of
$D(K_{Gr(2,N)})$ and their representations as GLSM B-branes. We will consider
generators that are compactly supported, so they are in correspondence with
generators of $D(Gr(2,N))$.
In \cite{Lerche:2001vj}, two sets of generators have been constructed, and they
are termed R-basis and S-basis. We choose the so-called R-basis which
corresponds to Kapranov's exceptional collection for the Grassmannian
\cite{Kapranov1988} (when pushed forward to $Gr(2,N)$). These generators
correspond to the sheaves
\begin{equation}\label{Rcomplex}
	i_*(L_\mu S) \cong \mathcal{C} \otimes \pi^*(L_\mu S)\in D(K_{Gr(2,N)}),
\end{equation}
where $i: Gr(2,N) \rightarrow K_{Gr(2,N)}$ is the embedding, $\pi$ is the projection map $\pi: K_{Gr(2,N)} \rightarrow Gr(2,N)$,
$\mathcal{C}$ is the complex
\begin{equation}
\mathcal{C}:=\pi^*(\det S)^{\otimes N} \rightarrow \mathcal{O},
\end{equation}
and $\mu$ is taken to be the young tableaux restricted to the rectangle:
\begin{equation}
	\{\mu \; | \;0\leq \mu_1 \leq N-k, \mu_i=0 \text{ for } i>k\}.
\end{equation}
The corresponding GLSM B-brane is straightforwardly constructed as reviewed in
Sec. \ref{sec:BbranesonGr}. As an example, we present the generators for
\begin{eqnarray}
 K_{Gr(2,4)}: \qquad \mu &\in&
\{(0,0), (1,0), (1,1), (2,0), (2,1), (2,2)\},\nonumber\\
K_{Gr(2,5)}:\qquad
\mu &\in& \{(0,0), (1,0), (1,1), (2,0), (2,1), (2,2), (3,0), (3,1), (3,2),
(3,3)\}.\nonumber\\
\end{eqnarray}
In their above form, none of the branes satisfy the grade restriction rules
listed in Eq. (\ref{GRR2a})-(\ref{GRR2c}). This is due to the fact that
the gauge charges of the complex $\mathcal{C}$, or any of its twists, always
violate the inequalities (\ref{GRR2a})-(\ref{GRR2c}). Therefore, the diagonal
`length' of the collection of gauge charges of the UV lift of $\mathcal{C} \otimes \pi^*(L_\mu S)$ is always, at least, $N$, which is too `wide' to satisfy \eqref{GRR2a}.

For example, take the case of $K_{Gr(2,5)}$ and Young tableau $\mu = (1,1)$. The brane is given by
\begin{align}
	\left( \pi^*(\det S)^{\otimes 5} \rightarrow \mathcal{O} \right) \otimes \pi^*\det S \cong i_* (\det S),
\end{align}
which is the IR image of the GLSM brane
\begin{equation}\label{examplebrane}
\mathcal{W}_{(6,6)} \rightarrow \mathcal{W}_{(1,1)}
\end{equation}
with the map given by the $P$ field.
The gauge charges of \eqref{examplebrane} are $(6,6)$ and $(1,1)$, so it is clear that
by twisting this brane we cannot restrict its gauge charges to satisfy
(\ref{GRR2a})-(\ref{GRR2c}). In order to grade restrict a brane, i.e. bring
all its charges inside a charge window, we need to replace certain factors of
the complex by binding with an empty brane. We will review this process in the next subsection.

\subsection{Grade restricting branes}

The process of grade restricting consists of bringing an arbitrary brane to an
equivalent one that satisfies the inequalities (\ref{GRR2a})-(\ref{GRR2c}) for
some fixed value of $\theta$. This is performed by binding the branes with an
`empty brane', i.e. a brane that RG-flows to a null-homotopic object in the IR at a fixed value of $t$. Mathematically, binding corresponds to taking
cones of an appropriate morphism between the branes in question. There are two
important properties we must keep in mind though:
\begin{enumerate}
 \item For a brane $\mathcal{E}$, being empty depends on the value of $t$
(more precisely, it depends on which chamber or phase $t$ belongs to). In our
examples, it depends on if we are considering $\xi\gg 1$ or $\xi\ll -1$.
\item If we have a brane $\mathcal{B}$ and an empty brane $\mathcal{E}$, in a
given phase, then
$$
Z_{D^{2}}(\mathcal{B})=Z_{D^{2}}(\mathrm{Cone}(\mathcal{E}\rightarrow
\mathcal{B})),
$$
when computed in such a phase.
\end{enumerate}
For a more detailed explanation and physical meaning of empty branes and
their role, see \cite{Herbst:2008jq,Hori:2013ika,hori2019notes}.
Once a brane is grade restricted, the resulting
brane is IR-equivalent to the original and by virtue of its gauge charges
fitting into (\ref{GRR2a})-(\ref{GRR2c}) it can be transported to the other
phase in the sense of \cite{Herbst:2008jq}. Therefore finding the empty
branes is a crucial step.

\subsubsection{Generating empty branes}

Let us first focus on the empty branes on the $\xi \gg 1$ phase. In this
phase, empty branes correspond to exact sequences of coherent sheaves. We can construct exact
sequences by using the results in
\cite{Donovan:2013gia,Donovan_2014}, which gives a systematic construction of
exact sequences in $Gr(k,N)$ as free resolutions of certain bundles. Then these
can be straightforwardly lifted to $K_{Gr(k,N)}$. For example, for $Gr(2,4)$
we have the following exact sequences\footnote{These complexes are presented as
objects of $D(Gr(2,N))$, but is clear that we can use them to grade restrict
objects in $D(K_{Gr(k,N)})$, as it is done in \cite{EHKR,Eager2017}.}:
\begin{align}
	S^{\vee(3,2)} \rightarrow S^{\vee(2,2) \; \oplus 4} \rightarrow S^{\vee(1,1) \; \oplus 4} \rightarrow S^{\vee}, \\
	S^{\vee(3,3)} \rightarrow S^{\vee(2,2) \; \oplus 6} \rightarrow S^{\vee(2,1) \; \oplus 4} \rightarrow S^{\vee(2)},
\end{align}
where we have defined
\begin{equation}
\mathcal{E}^{\mu} := L_\mu \mathcal{E}
\end{equation}
for any vector bundle $\mathcal{E}$.

Another way to produce exact sequences that will prove useful will be to use the fact that the Euler
sequence on $Gr(2,N)$
\begin{align}
	 S \rightarrow V \rightarrow Q,
\end{align}
where $Q$ denotes the rank $N-2$ universal quotient bundle on $Gr(2,N)$ and $V$
is the rank
$N$ trivial bundle, implies that the following is an exact
sequence\footnote{This is the content of Exercise 21 in Ch.2 of
\cite{weyman_2003}. However we remark that the functor $L_{\mu}$ in this paper corresponds to $L_{\mu^{T}}$ (denoted $L_{\mu'}$) in \cite{weyman_2003}, where $T$ denotes the transpose.}
\begin{align}
	 L_{\mu}[S \rightarrow V] \rightarrow L_{\mu}Q.
\end{align}
However the objects $L_{\mu}Q$ cannot be directly lifted to GLSM branes using
$U(2)$ representations. It requires the use of quasi-isomorphic objects. More
precisely, we can always express
\begin{align}
	L_{\mu}Q\cong \text{det}^{|\lambda|} S^\vee \otimes
L_{\lambda} Q^{\vee}\cong \text{det}^{|\lambda|} S^\vee \otimes
L_{\lambda^{T}}[V^\vee \rightarrow S^\vee ],
\end{align}
where $\lambda$ is an appropriate Young tableau that can be determined by basic bundle identities. We present an example in
detail in Appendix \ref{EmptGen}. Here, $\lambda^{T}$ denotes the transpose of
$\lambda$ and $\cong$ denotes quasi-isomorphisms.

The object $L_{\mu}[S \rightarrow V]$ is quasi-isomorphic to the long sequence:
\begin{equation}
	L_{\mu} [S \rightarrow V] \cong \epsilon_r^{\mu} \rightarrow
\epsilon_{r-1}^{\mu} \rightarrow ... \rightarrow \epsilon_1^{\mu}
\rightarrow \epsilon_0^{\mu},
\end{equation}
where $r$ is the number of boxes in the Young diagram $\mu$, and the factors of
the sequence are given by \cite{weyman_2003}:
\begin{equation}
	\epsilon_t^{\mu} = \bigoplus_{|\nu|=r-t} \left( \bigoplus_{|\omega|=t}
\mathbb{C}^{C^{\mu^{T}}_{\omega \nu}}\otimes L_\omega S \right) \bigotimes
L_{\nu^T} V.
\end{equation}
Here $|\nu|$ is the number of boxes of the young diagram $\nu$ and $C^{\mu^{T}}_{\omega \nu}$
are the Littlewood-Richardson coefficients.

The following two exact sequences are examples found via this method for $Gr(2,4)$. Note that these are resolutions of bundles corresponding to varying representations of $U(2)$, and the copies of the trivial bundles arise from Schur functors acting on $V$.
\begin{align}
	S \rightarrow \mathcal{O}^{\oplus 4} \rightarrow & S^{\vee(1,1) \; \oplus 4} \rightarrow S^{\vee(2,1)}, \label{G24Swap} \\
	S^{(1,1)} \rightarrow S^{\oplus 4} \rightarrow \mathcal{O}^{\oplus 10} \rightarrow & S^{\vee(2,2) \; \oplus 10} \rightarrow S^{\vee(3,2) \; \oplus 4} \rightarrow S^{\vee(3,3)}.
\end{align}

\subsubsection{An example of grade restriction in the $\xi\gg1$ phase}

In order to illustrate the process of grade restriction, we present an
example in this subsection. Consider the GLSM for $K_{Gr(2,5)}$. Then
denote the set of restricted charges, obtained from
(\ref{GRR2a})-(\ref{GRR2c}) corresponding to $-\pi < \theta < 0$, by
$\omega_{-1}$ in Fig. \ref{CharWin25-2}, and the brane given by the complex
\begin{equation}\label{sample25}
	\mathcal{W}_{(3,2)}  \rightarrow \mathcal{W}_{(-2,-3)},
\end{equation}
whose IR image in the geometric phase is $i_* S^{\vee(3,2)}$.
The weights/gauge charges carried by \eqref{sample25} are
$\{(-3,-2), (-2,-3)\}$ and $\{(3,2), (2,3)\}$. Then,
$\mathcal{W}_{(3,2)}\in\omega_{-1}$, but $\mathcal{W}_{(-2,-3)} \notin \omega_{-1}$ and we are instructed to replace it using the empty branes. This is done by binding the
brane with (i.e. by taking cones over) an empty brane on the $\xi\gg 1$ phase,
and then using brane-antibrane annihilation, i.e. removing the subcomplexes of
the form $\mathcal{W}\stackrel{\mathrm{Id}}{\rightarrow} \mathcal{W}$, to cast off the offending module.

Consider then the empty brane on $\xi\gg 1$ given by a twist of the
resolution of $S^{\vee(2)}$:
\begin{equation}\label{empty25s}
	S^{\vee(3,2)} \rightarrow S^{\vee(2,2) \oplus 5} \rightarrow S^{\vee(1,1) \oplus 10} \rightarrow  S^{\vee \oplus 5} \rightarrow  S^{\vee(2)} \otimes S^{(1,1)}.
\end{equation}
Binding the brane \eqref{sample25} and the UV lift of \eqref{empty25s} gives
\begin{align}\label{25cone1}
	\begin{array}{cccccccccc}
		& \cancel{\mathcal{W}_{(-2,-3)}} & \rightarrow & \mathcal{W}^{\oplus 5}_{(-2,-2)} & \rightarrow & \mathcal{W}_{(-1,-1)}^{\oplus 10} & \rightarrow & \mathcal{W}_{(0,-1)}^{\oplus 5} & \rightarrow & \mathcal{W}_{(1,-1)}\\
		& \oplus & \searrow & \oplus & & & & & & \\
		& \mathcal{W}_{(3,2)} & \rightarrow & \cancel{\mathcal{W}_{(-2,-3)}} & & & & & &
	\end{array},
\end{align}
where the crossed-out modules annihilate each other via brane-antibrane annihilation.
The cone with this empty brane has introduced another factor
$\mathcal{W}_{(1,-1)}\notin \omega_{-1}$ in the resulting complex,
since its charges include the weights $(1,-1)$ and $(-1,1)$ lying outside the
charge window. This factor can be
removed with the help of another exact sequence, which is the dual of \eqref{empty25s}:
\begin{align}\label{dualempty25s}
	S^{(2)} \otimes S^{\vee(1,1)} \rightarrow S^{\oplus 5} \rightarrow S^{(1,1) \oplus 10} \rightarrow  S^{(2,2) \oplus 5} \rightarrow  S^{(3,2)}.
\end{align}
Binding the UV lift of \eqref{dualempty25s} to the brane \eqref{25cone1} gives
\begin{align}
	\begin{array}{c@{}c@{}c@{}c@{}c@{}c@{}c@{}c@{}c@{}c@{}c@{}c@{}c@{}c@{}c@{}}
		 &  &  &  &  &  & \cancel{\mathcal{W}_{(1,-1)}} & \rightarrow & \mathcal{W}_{(1,0)}^{\oplus 5} & \rightarrow & \mathcal{W}^{\oplus 10}_{(1,1)} & \rightarrow & \mathcal{W}_{(2,2)}^{\oplus 5} & \rightarrow & \mathcal{W}_{(3,2)} \\
		& &  & & & & \oplus & \searrow & \oplus & & & & & &  \\
	\mathcal{W}_{(3,2)}	& \rightarrow & \mathcal{W}^{\oplus 5}_{(-2,-2)} & \rightarrow & \mathcal{W}^{\oplus 10}_{(-1,-1)} & \rightarrow & \mathcal{W}^{\oplus 5}_{(0,-1)} & \rightarrow & \cancel{\mathcal{W}_{(1,-1)}} &  &  &  &  &  & 
	\end{array}.
\end{align}

After brane-antibrane annihilation takes place, the resulting brane is grade restricted:
\begin{align}\label{restrictedex}
	\begin{array}{c@{}c@{}c@{}c@{}c@{}c@{}c@{}c@{}c@{}c@{}c@{}c@{}c@{}c@{}c@{}}
		\mathcal{W}_{(3,2)} & \rightarrow & \mathcal{W}_{(-2,-2)}^{\oplus 5} & \rightarrow & \mathcal{W}_{(-1,-1)}^{\oplus 10} & \rightarrow & \mathcal{W}^{\oplus 5}_{(0,-1)} & \rightarrow & \mathcal{W}^{\oplus 5}_{(1,0)} & \rightarrow & \mathcal{W}^{\oplus 10}_{(1,1)} & \rightarrow & \mathcal{W}^{\oplus 5}_{(2,2)} & \rightarrow & \mathcal{W}_{(3,2)}
	\end{array}.
\end{align}
The set of all weights presented in \eqref{restrictedex} fit
into $\omega_{-1}$.

All branes in $MF_{G}(W)$ can be grade restricted according to this
procedure. The end
result is an IR-equivalent brane, in the $\xi \gg 1$ (or, in the phase of
choice) that can `pass through the phase boundary smoothly' along a path within
the chosen range of the theta-angle on the moduli space, i.e. we can find an
admissible contour $L_{t}$ for every point on the path \cite{Hori:2006dk}.

\subsection{The window category for straight paths}\label{sec:evenresol}
We have seen in Sec. \ref{sec:GRR} that Eqs. (\ref{GRR2a})-(\ref{GRR2c})
guarantee the convergence of the hemisphere partition function along the phase
boundary. The objects $\mathcal{B}$ in $MF_{G}(W)$ satisfying
(\ref{GRR2a})-(\ref{GRR2c}) are expected to form a subcategory
$\omega_{l}\subset MF_{G}(W)$ where $l$ is an appropriate label associated with
the choice of $\theta$, we fix it to $l=\lfloor\frac{\theta}{\pi}\rfloor$. We
will refer to the categories $\omega_{l}$ as window categories\footnote{We
will abuse notation sometimes to use the same symbol $\omega_{l}$ to refer to the
subset of weights of $\rho_{M}$ that defines the window category.} and it is
expected that
\begin{align}\label{equivLV}
\omega_{l}\cong D(K_{Gr(k,N)})
\end{align}
for any $l$.

\begin{figure}[h]
	\centering
	\includegraphics[width=0.45 \linewidth]{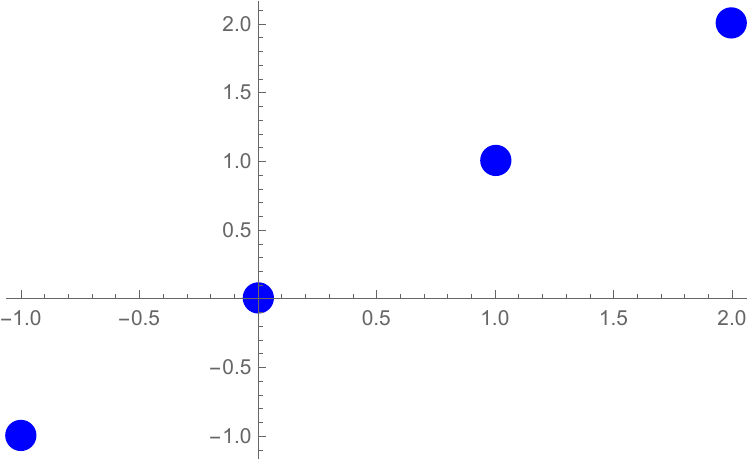}
	\caption{The strict charge window of $ -2\pi < \theta < 0 $ in the case of $K_{Gr(2,4)}$.}
	\label{CharWin24Strict}
\end{figure}

For $k=2$ and $N$ odd, the window categories can be found to be equivalent to Kuznetsov's exceptional collection\cite{Kuznetsov_2008}. This can be shown as
follows: from the relation
between the charges and the vector bundles on $Gr(2,N)$. There is a one-to-one
correspondence between the charges and the weights of $U(2)$, so they can be
assembled together to form various representations of $U(2)$. The charge $(m,m)$
corresponds to the $\det^m$ representation and therefore gives rise to $\pi^*\det^m
S$ (or $\pi^*\det^{\abs{m}} S^\vee$ when $m$ is negative), charges $\{ (n-i, i) |
i=0,1,\cdots, n \}$ correspond to the $\mathrm{Sym}^n$ representation and
therefore give rise to $\pi^*\mathrm{Sym}^n S$. It is then easy to check that, for a
suitably-chosen $\theta$ such that $(0,0)$ is in the window and all the charge
components are nonpositive, the representations correspond exactly to
Kuznetsov's exceptional collection\cite{Kuznetsov_2008}. Indeed, by writing $N=2n+1$ and defining $s:=\lfloor\frac{l+1}{2}\rfloor$ for $\theta\in(l\pi,\pi(l+1))$, the corresponding window $\omega_{l}$ is given by:
\begin{enumerate}
 \item For $l$ even, we define $\delta_{r}:=0$ if $r$ is even and $\delta_{r}:=1$ if $r$ is odd. Then
 \begin{equation}
\omega_{l}=\left\{\mathcal{W}_{(r+j-\lfloor\frac{r}{2}\rfloor-\delta_{r},~j-\lfloor\frac{r}{2}\rfloor-\delta_{r})}:-n\leq j\leq n,0\leq r\leq n-1\right\}\otimes \mathcal{W}_{(-s,-s)},
\end{equation}
whose IR image in the geometric phase is
 \begin{equation}
\pi^*\left(\left\{S^{(r)}\otimes\mathrm{det}^{j-\lfloor\frac{r}{2}\rfloor-\delta_{r}}S:-n\leq j\leq n,0\leq r\leq n-1\right\}\otimes\mathrm{det}^{-s}S \right).
\end{equation}
\item For $l$ odd
\begin{equation}
\omega_{l} = \left\{\mathcal{W}_{(r+j-\lfloor\frac{r}{2}\rfloor,~j-\lfloor\frac{r}{2}\rfloor)}:-n\leq j\leq n,0\leq r\leq n-1\right\}\otimes \mathcal{W}_{(-s,-s)},
\end{equation}
whose IR image in the geometric phase is
 \begin{equation}
\pi^*\left( \left\{S^{(r)}\otimes\mathrm{det}^{j-\lfloor\frac{r}{2}\rfloor}S:-n\leq j\leq n,0\leq r\leq n-1\right\}\otimes\mathrm{det}^{-s}S \right).
\end{equation}
\end{enumerate}
However, for $k=2$ and even $N$, we found a mismatch, i.e. \eqref{equivLV} does not hold\footnote{We remark that $U(2)$ GLSMs with an even number of chiral matter in the fundamental representation have been analyzed in \cite{Hori:2006dk}, finding additional difficulties, correlated with the existence of an irregular phase, as defined in \cite{Hori:2011pd}.}. Consider the case of $K_{Gr(2,4)}$. Note that in this case (and only in this case) we identify
\begin{align}
\omega_{2l}\equiv \omega_{2l+1},\qquad \text{ \ for \ }N=4\text{\ and all\ }l\in\mathbb{Z}
\end{align}
because there is no singularity at $\theta=\pi$.
The
gauge charges satisfying (\ref{GRR2a})-(\ref{GRR2c}) only contain line
bundles, making it impossible to fully generate $D(K_{Gr(2,4)})$. For example
for  $ -2\pi < \theta < 0 $, i.e. for $\omega_{-2}$, we plot the window
weights in Fig. \ref{CharWin24Strict}. Then, any $\omega_{s}$ only
contains a subcategory generated by four line bundles. This mismatch is caused
by the condition \eqref{GRR2b}, which in this case restricts the charges to the
diagonal: $\left|q^1 - q^2 \right| < 1$, that is the branes must only
support powers of the determinant representation.

\begin{figure}[h]
	\centering
	\includegraphics[width=0.45 \linewidth]{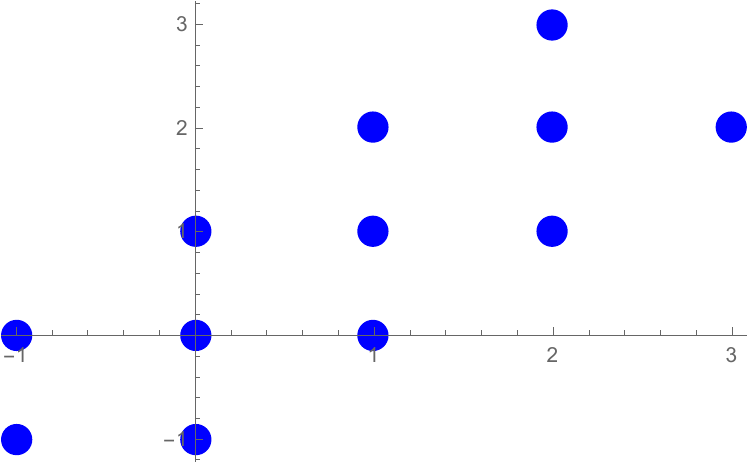}
	\caption{The relaxed charge window of $ -2\pi < \theta < 0 $ in the case of $K_{Gr(2,4)}$.}
	\label{CharWin24Relax}
\end{figure}

On the other hand, we can relax the conditions (\ref{GRR2a})-(\ref{GRR2c}) by allowing equalities instead of strict inequalities, only for the case of
$N$ even. The new conditions then read
\begin{align}
	\left| \frac{\theta}{\pi} + q^1 + q^2 \right| \leq N ;\label{GRR2aRelax} \\
	\left|q^1 - q^2 \right| \leq \frac{1}{2}(N-2); \label{GRR2bRelax} \\
	\left| \frac{\theta}{2 \pi} + q^a \right| \leq \frac{1}{4}(3N-2). \label{GRR2cRelax}
\end{align}
We can interpret the equality sign as allowing the exponential in Eq.
(\ref{part1}) to converge to a constant as $|\sigma| \rightarrow \infty$, and so
one hopes for a conditional convergence of the integral in this case, instead
of the strongest condition of absolute convergence. However, this cannot be the
case for all charges: Fig. \ref{CharWin24Relax} shows $\omega_{-2}$ for
(\ref{GRR2aRelax})-(\ref{GRR2cRelax}) and then we have the opposite
problem: there are now too many bundles -- eight in total -- so, it seems we
have a redundant description. Mathematically replacing $\omega_{-2}$ by
(\ref{GRR2aRelax})-(\ref{GRR2cRelax}) will be a solution for satisfying
\eqref{equivLV}. However, we run into a contradiction with the IR dynamics.

First, the restrictions (\ref{GRR2aRelax})-(\ref{GRR2cRelax}) predict a change
in the charge windows as one tunes $\theta$ across $\pi$, implying that
there is a singularity present, which is not the case for $K_{Gr(2,4)}$: a
singularity is only present at $\theta=0$.


Second, if we take the exact sequence
\eqref{G24Swap} and twist by $(\det S)^{\otimes 2}$, we get the new empty brane
\begin{equation}\label{emptybraneA}
\mathcal{W}_{(3,2)} \rightarrow \mathcal{W}_{(2,2)}^{\oplus 4} \rightarrow  \mathcal{W}_{(1,1)}^{\oplus 4} \rightarrow \mathcal{W}_{(1,0)}\in\hat{\omega}_{-2}.
\end{equation}
then, this empty brane exacerbates the problem: the set of its charges fit
snugly into the charge window of Fig. \ref{CharWin24Relax}. This is quite
troublesome for an empty brane: such a charge window suggests that it,
originally empty in the $\xi\gg 1$ phase, can transport safely to the $\xi\ll
-1$ phase where it becomes nonempty, leading to a contradiction. We interpret
this as \eqref{emptybraneA} imposing relations among the branes in the charge
window of Fig. \ref{CharWin24Relax}. Let us explain this more clearly, denote
the window category defined by (\ref{GRR2aRelax})-(\ref{GRR2cRelax}) as
\begin{equation}
\hat{\omega}_{s}:\{\text{weights of}~\rho_{M}~\text{satisfying~(\ref{GRR2aRelax})-(\ref{GRR2cRelax})}\}
\end{equation}

\begin{figure}[h]
	\begin{subfigure}[b]{0.35 \linewidth}
		\includegraphics[width=\linewidth]{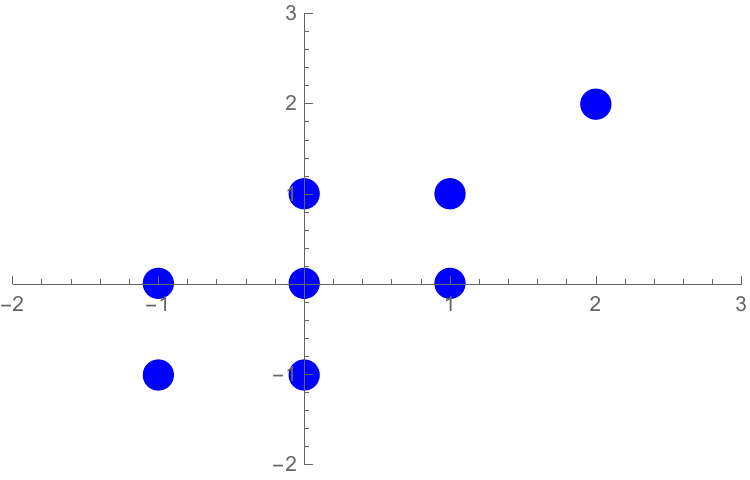}
		\caption{Type 1}
		\label{G24Type1}
	\end{subfigure}
	\begin{subfigure}[b]{0.35 \linewidth}
		\includegraphics[width=\linewidth]{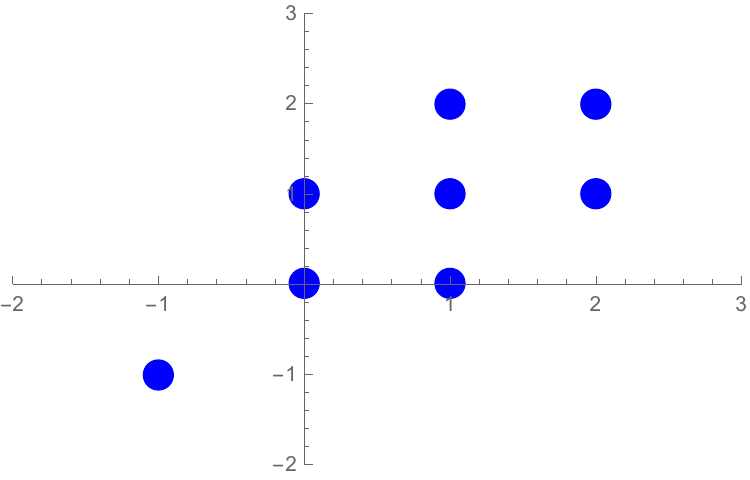}
		\caption{Type 2}
		\label{G24Type2}
	\end{subfigure}
	\centering
	\caption{Two possible charge windows for $K_{Gr(2,4)}$, $-2 \pi < \theta <
0$.}
	\label{G24MonWins2}
\end{figure}

Take a brane $\mathcal{B}\in\hat{\omega}_{-2}$. Then, we have two empty
branes that fit into $\hat{\omega}_{-2}$, one is \eqref{emptybraneA}, but we
can also twist it by $\mathcal{W}_{(-1,-1)}$ and it will still fit into $\hat{\omega}_{-2}$:
\begin{equation}\label{emptybraneA2}
\mathcal{W}_{(2,1)} \rightarrow \mathcal{W}_{(1,1)}^{\oplus 4} \rightarrow  \mathcal{O}^{\oplus 4} \rightarrow \mathcal{W}_{(0,-1)}\in\hat{\omega}_{-2}.
\end{equation}
We can then bind enough copies of \eqref{emptybraneA} and
\eqref{emptybraneA2} to $\mathcal{B}$ and annihilate all occurrences of any
pair of branes, one from the set $\{\mathcal{W}_{(3,2)},\mathcal{W}_{(1,0)}\}$ and one from the set $\{\mathcal{W}_{(0,-1)},\mathcal{W}_{(2,1)}\}$. For example, if we choose to annihilate all instances of the modules $\mathcal{W}_{(2,1)}$ and $\mathcal{W}_{(3,2)}$, we can whittle down $\hat{\omega}_{-2}$ to the window illustrated in Fig. \ref{G24Type1}. On the
other hand if we choose to annihilate $\mathcal{W}_{(0,-1)}$ and $\mathcal{W}_{(3,2)}$, this would result in the charge diagram of Fig. \ref{G24Type2}. There are two
other possibilities, but they are disregarded as they can be shown to
vary across $\theta = \pi$, where no singularity is present. In summary, the possibilities are, for subwindows of $\hat{\omega}_{2l}$ (for $\theta\in(2\pi l,2\pi(l+1))$, $l\in\mathbb{Z}$) with the shape as in Fig. \ref{G24Type1}, given by
 \begin{eqnarray}
\omega^{(1)}_{2l}&:=&\{\mathcal{W}_{(-2,-2)},\mathcal{W}_{(-1,-1)},\mathcal{W}_{(0,0)},\mathcal{W}_{(1,1)},\mathcal{W}_{(0,-1)},\mathcal{W}_{(-1,-2)}\}\otimes \mathcal{W}_{(-l,-l)}.
\end{eqnarray}
On the other hand, the windows with the shape as in Fig. \ref{G24Type2} are given by
\begin{eqnarray}
\omega^{(2)}_{2l}&:=&\{\mathcal{W}_{(-2,-2)},\mathcal{W}_{(-1,-1)},\mathcal{W}_{(0,0)},\mathcal{W}_{(1,1)},\mathcal{W}_{(1,0)},\mathcal{W}_{(0,-1)}\}\otimes \mathcal{W}_{(-l,-l)}.
\end{eqnarray}

In Sec. \ref{sec:open}, by computing the open Witten indices, we will
argue that both choices for the charge restrictions appear to be equally valid
and there is no reason to prefer one over the other. However, when considering
monodromy the situation is a bit more involved.

For $N>4$, there are singular points at both $\theta=0$ and $\theta=\pi$, so
none of the possible whittled-down charge restrictions can be discarded on the
grounds of presenting a discontinuity at $\theta=0,\pi$ as we did in the case
$N=4$. The general pattern for the charge window of $K_{Gr(2,2n)}$ for $n\in
\mathbb{Z}_{> 2}$ is given by any possible subset of
$\hat{\omega}_{l}$, defined by (\ref{GRR2aRelax})-(\ref{GRR2cRelax}), and then
we annihilate any choice of $n$ branes by binding with the ($\xi\gg 1$) empty
branes obtained from the exact sequences
\begin{eqnarray}\label{eeemptybrane}
&&\mathrm{det} S^{\vee}\otimes(V^{(0)}\otimes
S^{\vee(n-1)}\rightarrow V^{(1)}\otimes
S^{\vee(n-2)}\rightarrow\cdots\rightarrow V^{(n-1)}\otimes
S^{\vee(0)})\rightarrow\nonumber\\
&&\rightarrow V^{(n-1)}\otimes
S^{(0)}\rightarrow V^{(n-2)}\otimes
S^{(1)}\rightarrow\cdots\rightarrow V^{(0)}\otimes
S^{(n-1)}
\end{eqnarray}
or any of its $n-1$ twists that remain in $\hat{\omega}_{l}$. This gives us a
choice of $2^n$ possible sub-windows of $\hat{\omega}_{l}$ (for each $l$). A
simple description of these choices can be described as follows: write $N=2n$ and consider the points $q^{1}-q^{2}=\pm(n-1)$ in
$\hat{\omega}_{l}$. Each sign choice corresponds to the `outermost diagonal' of the charge window $\hat{\omega}_{l}$ and each outermost diagonal has $2n$ points. If in addition we restrict the constraint (\ref{GRR2aRelax}) a bit more to
\begin{align}\label{extraconditions}
	\left| \frac{\theta}{\pi} + q^1 + q^2 \right| \leq n,\qquad \text{ \ for \ }q^{1}-q^{2}=\pm(n-1),
\end{align}
then we can define a new window $\omega'_{l}\subset\hat{\omega}_{l}$ by
\begin{align}\label{primewindoww}
	\omega'_{l}:=\left\{\text{weights of}~\rho_{M}~\text{satisfying}~(\ref{GRR2aRelax})~\text{and}~(\ref{extraconditions})\right\}\subset \hat{\omega}_{l}.
\end{align}
We claim it spans an exceptional collection for $D(K_{Gr(2,N)})$. Indeed, explicitly, we can write, for $\theta\in(l\pi,\pi(l+1))$ and $s:=\lfloor\frac{l+1}{2}\rfloor$, the corresponding window $\omega'_{l}$ is given by:
\begin{enumerate}
 \item For $l$ even
\begin{eqnarray}
\omega'_{l}&=&\left\{\mathcal{W}_{(r+j-\lfloor\frac{r}{2}\rfloor,~j-\lfloor\frac{r}{2}\rfloor)}:-n\leq j\leq n-1,0\leq r\leq n-2\right\}\otimes \mathcal{W}_{(-s,-s)} \nonumber\\
&\cup&\left\{\mathcal{W}_{(n-1+j-\lfloor\frac{n-1}{2}\rfloor,~j-\lfloor\frac{n-1}{2}\rfloor)}:-n\leq j\leq -1\right\}\otimes \mathcal{W}_{(-s,-s)},
\end{eqnarray}
whose IR image in the geometric phase is 
 \begin{eqnarray}
&&\pi^*\left(\left\{S^{(r)}\otimes\mathrm{det}^{j-\lfloor\frac{r}{2}\rfloor}S:-n\leq j\leq n-1,0\leq r\leq n-2\right\}\otimes\mathrm{det}^{-s}S\right)\nonumber\\
&\cup&\pi^*\left(\left\{S^{(n-1)}\otimes\mathrm{det}^{j-\lfloor\frac{n-1}{2}\rfloor}S:-n\leq j\leq -1\right\}\otimes\mathrm{det}^{-s}S\right).
\end{eqnarray}
\item For $l$ odd, we define $\delta_{r}:=0$ if $r$ is even and $\delta_{r}:=1$ if $r$ is odd. Then
\begin{eqnarray}
\omega'_{l}&=&\left\{\mathcal{W}_{(r+j-\lfloor\frac{r}{2}\rfloor-\delta_{r},~j-\lfloor\frac{r}{2}\rfloor-\delta_{r}}:-n+1\leq j\leq n,0\leq r\leq n-2\right\}\otimes \mathcal{W}_{(-s,-s)}
\nonumber\\
&\cup&\left\{\mathcal{W}_{(n-1+j-\lfloor\frac{n-1}{2}\rfloor-\delta_{n-1},~j-\lfloor\frac{n-1}{2}\rfloor-\delta_{n-1})}:-n+1\leq j\leq 0\right\}\otimes \mathcal{W}_{(-s,-s)},
\end{eqnarray}
whose IR image in the geometric phase is 
\begin{eqnarray}
&&\pi^*\left(\left\{S^{(r)}\otimes\mathrm{det}^{j-\lfloor\frac{r}{2}\rfloor-\delta_{r}}S:-n+1\leq j\leq n,0\leq r\leq n-2\right\}\otimes\mathrm{det}^{-s}S\right)
\nonumber\\
&\cup&\pi^*\left(\left\{S^{(n-1)}\otimes\mathrm{det}^{j-\lfloor\frac{n-1}{2}\rfloor-\delta_{n-1}}S:-n+1\leq j\leq 0\right\}\otimes\mathrm{det}^{-s}S\right).
\end{eqnarray}
\end{enumerate}
Note that $\omega'_{l}$ differs from $\hat{\omega}_{l}$ only in the outermost diagonals. Then all the other choices of valid windows can
be obtained by replacing points in $\omega'_{l}$ using the empty brane (\ref{eeemptybrane}) or any of its $n-1$ twists. This give us the possibility of choosing $2^n$ (and $2$ for $n=2$) different subwindows of $\hat{\omega}_{l}$ as our valid window. We remark that all these choices only differ by their points on the outermost diagonals.

\begin{figure}[h]
	\begin{subfigure}[b]{0.35 \linewidth}
		\includegraphics[width=\linewidth]{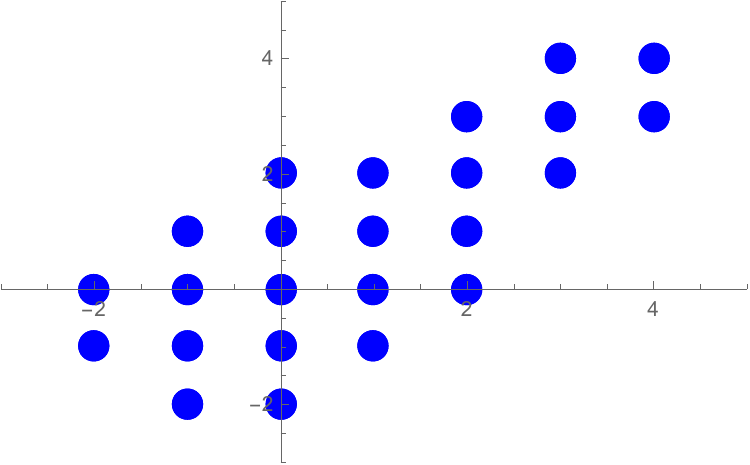}
	\end{subfigure}
	\begin{subfigure}[b]{0.35 \linewidth}
		\includegraphics[width=\linewidth]{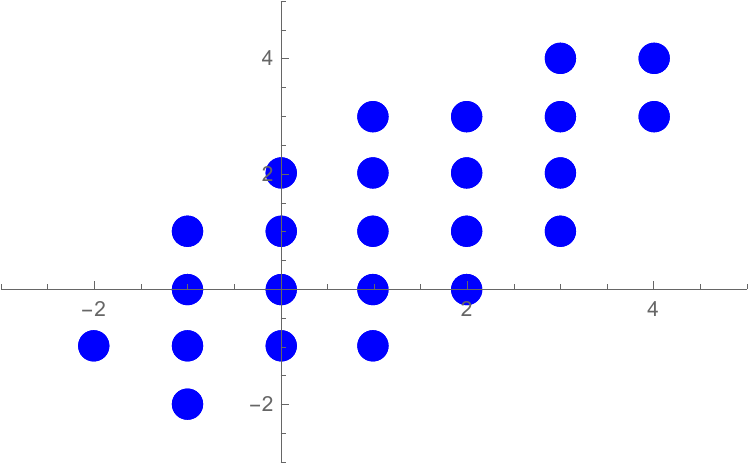}
	\end{subfigure}
	\centering
	
	\begin{subfigure}[b]{0.35 \linewidth}
		\includegraphics[width=\linewidth]{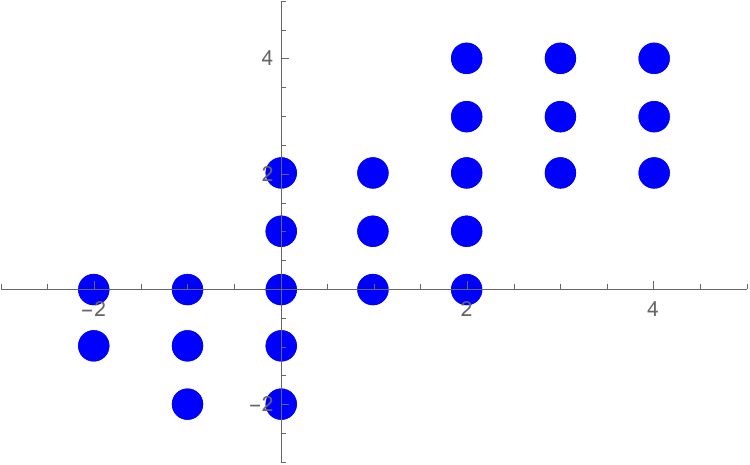}
	\end{subfigure}
	\begin{subfigure}[b]{0.35 \linewidth}
		\includegraphics[width=\linewidth]{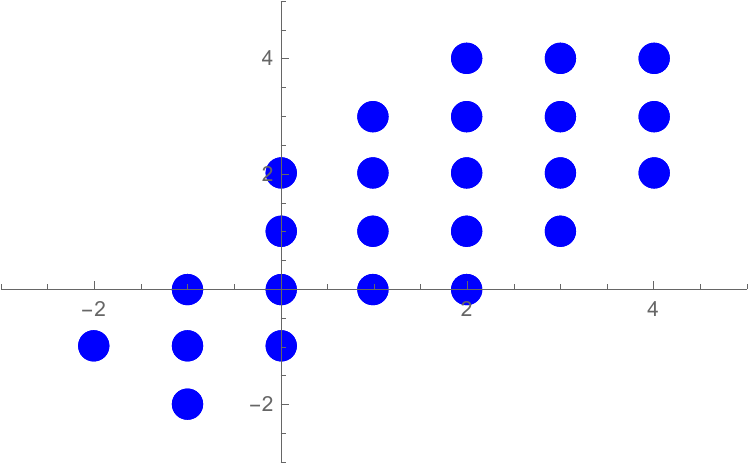}
	\end{subfigure}
	\centering
	\caption{A selection of valid charge windows $\omega_{-3}$ for $K_{Gr(2,6)}$ with $-3 \pi < \theta < -2 \pi$.}
	\label{G26PossWins}
\end{figure}

For example, see Fig. \ref{G26PossWins} for a selection of equally-valid charge
windows, all describing sub-windows of $\hat{\omega}_{-3}$ for $(k,N)=(2,6)$.
Therefore, we see that multiple charge windows are found to be acceptable for
each choice of $\theta\neq 0,\pi$ $\mathrm{mod}$ $2\pi$ in the even $N$ case.

In summary, the window category in the odd $N$ case is exactly given by Eq.
(\ref{GRR2a})-(\ref{GRR2c}). In the even $N$ case, we propose that we have
$2^{n}$ choices of sub-windows from $\hat{\omega}_{s}$ as described above,
except for $n=2$ where we have $2$ choices.

To transport a brane from one phase to the other through a window, any choice of
the generators for the window category is fine. We just need to fix one set of
generators and use the empty branes to grade restrict the branes to be
transported. However, if we want to transport the branes along a path encircling
a singular point, we need to consider two windows adjacent to each other, one to
the left and another to the right of the singular point. Then, as we will see in
Sec. \ref{sec:monodromy}, for $N$ even, if we choose one of the sub-windows,
the sub-window next to it is automatically fixed and can be distinct from the
first choice. This observation suggests that if a choice is made for one
sub-window, then all the others are fixed as long as
monodromy is taken into account.

\section{\label{sec:open}Open Witten index for
$K_{Gr(2,N)}$}
Up to now, we have presented how to determine the window categories and how to
grade restrict the branes. In this section we will present some piece of
evidence that our prescription for defining the window categories
$\omega_{s}$, in particular in the $N$ even case, is consistent. One way to
check is to calculate the open Witten index, a measure of the number of
supersymmetric ground states\cite{hori2000dbranes,hori2003mirror}, in the open
string sector. The
open Witten index is independent of the FI-theta parameter $t$, and hence is
expected to be identical when computed in any phase.

The open Witten index is computed by the partition function on an annulus
with boundary conditions specified by a pair of B-branes $\mathcal{B}_1,
\mathcal{B}_2$. The open Witten index is a topological quantity and can be
computed by the following localization formula on the Coulomb branch
\cite{Hori:2013ika}
\begin{equation}\label{wittenindex}
	\chi(\mathcal{B}_1,\mathcal{B}_2)=\frac{1}{|W_G|} \int_\Gamma \frac{d^{l_G}u}{(2 \pi i)^{l_G}} Z_{\text{Vec}}Z_{\text{Chi}} f_{\mathcal{B}_1}\left(-\frac{u}{2 \pi}\right) f_{\mathcal{B}_2}\left(\frac{u}{2 \pi}\right),
\end{equation}
where $W_G$ is the Weyl group of the gauge group $G$, $\Gamma \subset
\mathfrak{t}_{\mathbb{C}}/(2 \pi Q^{\vee})$ is a middle-dimensional contour in
the complexified Lie algebra of the maximal torus of $G$ $\mathrm{mod}$
$2 \pi Q^{\vee}$ ($Q^{\vee}$ denotes the dual weight lattice).
$f_{\mathcal{B}_j}$ is the brane factor of $\mathcal{B}_j$, as in Eq.
\eqref{Disk}, and the contributions from the vector multiplet and chiral
multiplets are given by
\begin{align}
	Z_{\text{Vec}} =& \prod_{\alpha>0} \left( 2 \sinh\left(
\frac{\alpha(u)}{2} \right) \right)^2, \\
	Z_{\text{Chi}} =& \frac{1}{\prod_i 2 \sinh\left(\frac{Q_i(u)}{2}
\right)}.
\end{align}

For $Z_{\text{Vec}}$, the product is over the positive roots of the Lie algebra. For $U(k)$, the roots are $k(k-1)$ permutations of
\begin{equation}
	\alpha_0=(1,-1,0,...,0).
\end{equation}
The positive roots are then chosen to be the $\frac{k}{2}(k-1)$ permutations where the first non-zero element is positive. This then gives
\begin{equation}
	Z_{\text{Vec}}^{k,N} = \prod_{i=1}^{k-1} \prod_{j>i}^k \left(2 \sinh \left(
\frac{1}{2}(u_i-u_j)\right) \right)^2.
\end{equation}

For the GLSM of $K_{Gr(k,N)}$ investigated here, $Z_{\text{Chi}}$ becomes
\begin{equation}
	\left( Z_{\text{Chi}}^{k,N} \right)^{-1}=2^{kN+1} \prod_{i=1}^k \sinh^N
\left(\frac{1}{2} u_i \right) \sinh\left(-\frac{N}{2} \sum_{i=1}^k
u_i\right).
\end{equation}

It is important to remark that the contour $\Gamma$ is left undetermined on
\cite{Hori:2013ika} and is only specified in a few examples. Here we do the same
and we will specify $\Gamma$ later on for our class of models.

The brane factors in Eq.\eqref{wittenindex} can be composed of Schur characters
of the representations at play. For the brane with IR
projection given by
\begin{align}
	i_* L_\mu S \cong \left( \pi^*(\det S)^{\otimes N} \rightarrow \mathcal{O} \right) \otimes \pi^*L_\mu S,
	\label{OrigBrane}
\end{align}
the brane factors are
\begin{equation}
	(1-e^{N\left(\sum_i^k u_i\right)}) f_\mu(e^{u_1},...,e^{u_k}),
\end{equation}
where $(1-e^{N\left(\sum_i^k u_i\right)})$ is the brane factor of the complex
$\mathcal{C} = \left( \pi^*(\det S)^{\otimes N} \rightarrow \mathcal{O} \right)$, and $f_\mu$, the brane factor of $\pi^*L_\mu S$, is the Schur polynomial associated with $\mu$.

For more general branes, $f_{\mathcal{B}}(u/2\pi)$ is determined
straightforwardly. For example, consider the $\xi\gg 1$ empty brane corresponding to the exact sequence on $Gr(2,4)$
\begin{align}
	S^{(2,1)} \rightarrow S^{(1,1) \; \oplus 4} \oplus S^{(2) \; \oplus 4} \rightarrow S^{\oplus 16} \rightarrow \mathcal{O}^{\oplus 20} \rightarrow S^{\vee(2,2) \; \oplus 4} \rightarrow S^{\vee(3,2)},
\end{align}
one can read off the brane factor by summing the Schur characters, with exponential variables, of the representations in each step of the sequence, with alternating signs, so the brane factor of the complex above can be written as:
\begin{align*}
	&\left( e^{2u_1 + u_2} + e^{u_1 + 2u_2} \right) - \left( 4 e^{u_1 + u_2 } + 4 (e^{2u_1} + e^{u_1 + u_2} + e^{2u_2}) \right) \\
	& + 16 \left( e^{u_1} + e^{u_2} \right) - 20\left( 1 \right) + 4\left( e^{-2u_1 - 2u_2} \right) - \left(e^{-3u_1 - 2u_2} + e^{-2u_1 - 3u_2} \right).	
\end{align*}

Putting everything together, for the $K_{Gr(k,N)}$ models, the open Witten
index becomes
\begin{equation}
	\chi(\mathcal{B}_1,\mathcal{B}_2) = \frac{1}{k!} \int_{\Gamma} \frac{d^k
u}{(2 \pi i)^k} \frac{\prod_{i=1}^{k-1} \prod_{j>i}^k \left(2 \sinh
\left( \frac{1}{2}(u_i-u_j)\right) \right)^2 f_{\mathcal{B}_1}\left(-\frac{u}{2
\pi}\right) f_{\mathcal{B}_2}\left(\frac{u}{2 \pi}\right) }{2^{kN+1}
\prod_{i=1}^k \sinh^N \left(\frac{1}{2} u_i \right)
\sinh\left(-\frac{N}{2} \sum_{i=1}^k u_i \right)}.
\end{equation}

One can evaluate the integral using the techniques for multidimensional
residues from Mellin-Barnes integrals \cite{zhdanov1998computation}, or, for
$k=2$, from a composite residue theorem. In the $\xi\gg 1$ phase, $\Gamma$ can
be chosen as a torus around the origin:
\begin{equation}
\Gamma_{+}=\{|u_{a}|=\varepsilon:a=1,\ldots k\}
\end{equation}
with $\varepsilon$ sufficiently small. For the negative phase, the contour is a
bit more complicated for general $k$, but for $k=2$, $\Gamma_{-}$ can be
reduced to the union of tori encircling the poles located at $(u_1,u_2) =
\left(0, \frac{2 \pi i v}{N}\right)$, $v \in \{1,...,N-1\}$, and their Weyl
images, that is
\begin{equation}
\Gamma_{-}=\bigcup_{v=1}^{N-1}\left\{|u_{1}|=\varepsilon,\left|u_{2}-\frac{2
\pi i v}{N}\right|=\varepsilon\right\}\cup \left\{|u_{2}|=\varepsilon,\left|u_{1}-\frac{2
\pi i v}{N}\right|=\varepsilon\right\}.
\end{equation}
We then expect that $\chi(\mathcal{B}_1,\mathcal{B}_2)$ is independent of the
choice of contour, $\Gamma_{\pm}$, when $\mathcal{B}_1$ and $\mathcal{B}_2$ are
grade restricted.
In the case of $K_{Gr(k,N)}$, we can choose a set of generators
$\{\mathcal{B}_1,\mathcal{B}_2,\cdots,\mathcal{B}_{N \choose k}\}$, for
instance the ones corresponding to $\mathcal{W}_{\mu}$ constructed from
Kapranov's exceptional collection for $Gr(k,N)$. Then, we compute the open Witten
indices
\begin{equation}
(\chi^{\pm}_{k,N})_{ij} := \chi(\mathcal{B}_i,\mathcal{B}_j)
\end{equation}
with the contours $\Gamma_{\pm}$.
It is found that, when grade restricting the branes of Eq. \eqref{OrigBrane} to a specific window, the open Witten index was consistent when calculated on the positive phase and on the negative phase, i.e. $\chi^{+}_{k,N}=\chi^{-}_{k,N}$, suggesting strongly that successful brane transport has taken place. The results of a few examples are presented below.

For $K_{Gr(2,4)}$, the basis of branes used corresponded to the weights $\{(0,0), (1,0), (1,1), (2,0),$ $(2,1), (2,2)\}$ and the open Witten indices were calculated to be
\begin{align}
	\chi^+_{2,4} = \chi^-_{2,4} = \left(
	\begin{array}{cccccc}
		0 & 4 & 6 & 10 & 20 & 20 \\
		-4 & 0 & 4 & 4 & 16 & 20 \\
		-6 & -4 & 0 & 0 & 4 & 6 \\
		-10 & -4 & 0 & 0 & 4 & 10 \\
		-20 & -16 & -4 & -4 & 0 & 4 \\
		-20 & -20 & -6 & -10 & -4 & 0 \\
	\end{array}
	\right).
\end{align}
A special note is needed here: this was found to be consistent for a set of branes all grade restricted to one of the charge window types in Fig. \ref{G24MonWins2}. On the other hand, if a set of branes grade restricted to different window types were mixed together, so that the entire set was not grade restricted to the same charge window type, then the open Witten indices were found to be inconsistent. Thus it is believed that branes must be grade restricted in exactly the same way when computing the open Witten index, despite multiple ways of charge restriction being valid for a single choice of theta-angle.

For $K_{Gr(2,5)}$, the basis of branes used corresponded to the weights $\{(0,0), (1,0), (1,1), (2,0),$ $(2,1), (2,2),$ $(3,0), (3,1), (3,2), (3,3)\}$. Then we found\footnote{We remark that (\ref{chi25}) was computed in \cite{Lerche:2001vj}. It corresponds to the open Witten index computed in what is called the `S-basis' in \cite{Lerche:2001vj}. More precisely, using the notation of \cite{Lerche:2001vj}: $(\chi_{-}^{-T}-\chi_{-}^{-1})=\chi^+_{2,5} = \chi^-_{2,5}$.}
\begin{align}\label{chi25}
	\chi^+_{2,5} = \chi^-_{2,5} = \left(
	\begin{array}{cccccccccc}
		0 & -5 & -10 & -15 & -40 & -50 & -35 & -105 & -175 & -175 \\
		5 & 0 & -5 & -5 & -25 & -40 & -15 & -75 & -155 & -175 \\
		10 & 5 & 0 & 0 & -5 & -10 & 0 & -15 & -40 & -50 \\
		15 & 5 & 0 & 0 & -5 & -15 & -5 & -25 & -75 & -105 \\
		40 & 25 & 5 & 5 & 0 & -5 & 0 & -5 & -25 & -40 \\
		50 & 40 & 10 & 15 & 5 & 0 & 0 & 0 & -5 & -10 \\
		35 & 15 & 0 & 5 & 0 & 0 & 0 & -5 & -15 & -35 \\
		105 & 75 & 15 & 25 & 5 & 0 & 5 & 0 & -5 & -15 \\
		175 & 155 & 40 & 75 & 25 & 5 & 15 & 5 & 0 & -5 \\
		175 & 175 & 50 & 105 & 40 & 10 & 35 & 15 & 5 & 0 \\
	\end{array}
	\right).
\end{align}

For $K_{Gr(2,6)}$, we have fifteen generators so, we just present the
computation for the subset \\$\{(0,0), (1,0),
(1,1), (2,1), (2,2), (3,2),$ $(3,3), (4,3), (4,4)\}$:
\begin{align}
	\chi^+_{2,6} = \chi^-_{2,6} = \left(
	\begin{array}{ccccccccc}
		0 & 6 & 15 & 70 & 105 & 420 & 490 & 1764 & 1764 \\
		-6 & 0 & 6 & 36 & 70 & 315 & 420 & 1624 & 1764 \\
		-15 & -6 & 0 & 6 & 15 & 70 & 105 & 420 & 490 \\
		-70 & -36 & -6 & 0 & 6 & 36 & 70 & 315 & 420 \\
		-105 & -70 & -15 & -6 & 0 & 6 & 15 & 70 & 105 \\
		-420 & -315 & -70 & -36 & -6 & 0 & 6 & 36 & 70 \\
		-490 & -420 & -105 & -70 & -15 & -6 & 0 & 6 & 15 \\
		-1764 & -1624 & -420 & -315 & -70 & -36 & -6 & 0 & 6 \\
		-1764 & -1764 & -490 & -420 & -105 & -70 & -15 & -6 & 0 \\
	\end{array}
	\right).
\end{align}

The above results suggest strongly that branes have been properly grade restricted to fit valid charge windows and have been transported across a phase boundary into another phase in a way that has not incurred any inconsistencies.

\section{\label{sec:monodromy}Monodromy}

In this section we will study the transport of B-branes around a loop
$\mathcal{L}$ in $\mathcal{M}_{K}$. We will consider only loops with base point near $\xi\gg 1$ and made of compositions of straight paths from $\xi\gg 1$ to $\xi\ll-1$ or their inverses. A B-brane undergoing this process will
return to a copy of itself acted on by a transformation $\phi_{\mathcal{L}}$ .
Such transformations depend only on the homotopy class of the path\footnote{In the following we will encounter that after fixing the homotopy class of $\mathcal{L}$, we will have several choices of window categories, however we propose that the functor $\phi_{\mathcal{L}}$ is independet of this choice and indeed only depends on the homotopy class of $\mathcal{L}$.}
$\mathcal{L}$. If the basepoint of $\mathcal{L}$ belongs to a specific phase,
when transporting a B-brane $\mathcal{B}$ along $\mathcal{L}$ we will have to
map $\mathcal{B}$ to an IR equivalent B-brane using appropriate empty branes in
order to find the image of $\mathcal{B}$ under the equivalences
$\omega_{s}\rightarrow\omega_{s\pm 1}$. The composition of all these
transformations will result in the transformation $\phi_{\mathcal{L}}$, which
can be identified with a functor that belongs to the group of autoequivalences of the category
of IR B-branes, in the phase where we have set the basepoint
\cite{Herbst:2008jq}. In this work we will be concerned by paths $\mathcal{L}$
constructed from straight paths and with base point at the $\xi\gg 1$ phase and encircling all the poles above either $\theta=0$ or $\theta=\pi$. We will not consider other monodromies such as loops around limit phase points.
We expect
\begin{align}
\phi_{\mathcal{L}}\in\mathrm{Aut}(D(K_{Gr(2,N)})).
\end{align}

In general, the paths described above will not cover all possible
monodromies for the $K_{Gr(2,N)}$ models since, by definition, they will
encircle all the singularities located at either $\theta=0$ or $\theta=\pi$,
but not individual ones.

Below we will present first some illustrative examples of such monodromies
and the functor $\phi_{\mathcal{L}}$. These examples will show the
difference between this computation for $N$ even and odd. For the case $N$
even, we will face new challenges, due to the multiple choices of windows and
how this will affect the monodromy computation. Since this is a novel situation
we encounter in B-brane dynamics of GLSMs, we will propose a prescription to
compute the monodromy and show that it passes several consistency checks. After
the examples we will discuss the general case in Sec.
\ref{sec:genmonaction}. In the particular examples $N=4$ and $N=5$, the monodromies given
by composing straight paths actually encircle only one singularity at a time.
This is not the case for $N>5$. Then, in each case (where $\mathcal{L}$ encircles a single singularity), it is expected that
$\phi_{\mathcal{L}}$ takes the form of a spherical twist
\cite{huybrechts2006fourier} and therefore, the expected monodromy
transformation of the hemisphere partition function is given by
\begin{align}
	Z^+(\mathcal{B}') =& Z^+(\mathcal{B}) - \chi(\mathcal{B},\mathcal{E})
Z^+(\mathcal{E}),\qquad \mathcal{B}':=\phi_{\mathcal{L}}(\mathcal{B})
	\label{MonZCheck}
\end{align}
where the superindex $+$ indicates that the partition function has been
computed in the $\xi\gg 1$ phase. The object $\mathcal{E}$ corresponds to the
spherical object defining the functor $\phi_{\mathcal{L}}$. We will identify
$\mathcal{E}$ for each loop in the $N=4,5$ cases.

\subsection{Example: $K_{Gr(2,4)}$}\label{G24Mon}

We will present a detailed computation for a clockwise
monodromy around the singularity at $\theta=0$. This process involves two
window categories: one at $0 < \theta < 2\pi$ denoted $\omega_0$, and $-2\pi <
\theta < 0$ denoted $\omega_{-2}$. We have two valid choices for each window, as
we discussed in Sec. \ref{sec:evenresol}.

\begin{figure}[h]
	\begin{subfigure}[b]{0.45 \linewidth}
		\includegraphics[width=\linewidth]{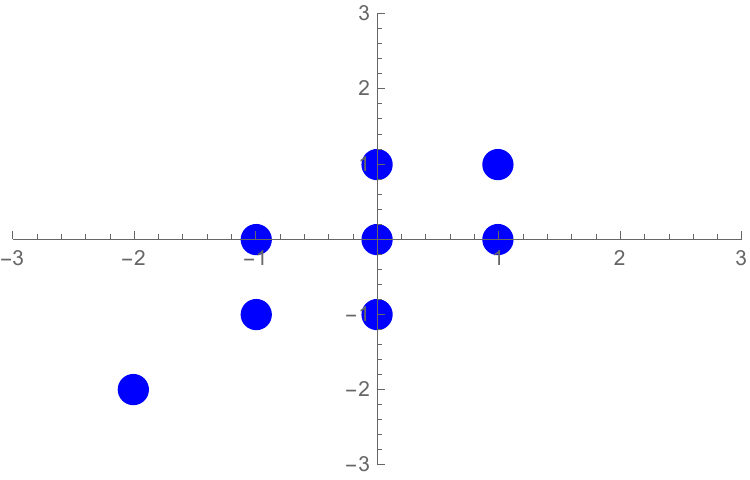}
		\caption{$\omega_{0}$}
		\label{G24Win0}
	\end{subfigure}
	\begin{subfigure}[b]{0.45 \linewidth}
		\includegraphics[width=\linewidth]{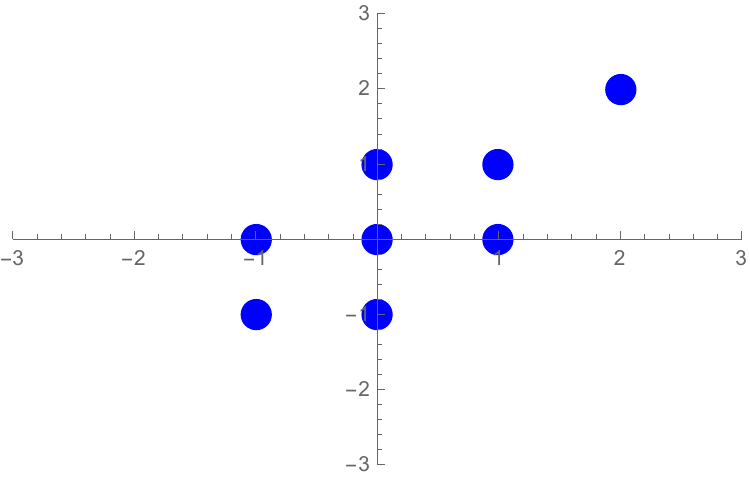}
		\caption{$\omega_{-2}$}
		\label{G24Win-2}
	\end{subfigure}
	\centering
	\caption{The charge restrictions for windows taken on the monodromy loop around $\theta=0$ of $K_{Gr(2,4)}$.}
	\label{G24CharRest}
\end{figure}

\begin{figure}[h]
	\begin{subfigure}[b]{0.45 \linewidth}
		\includegraphics[width=\linewidth]{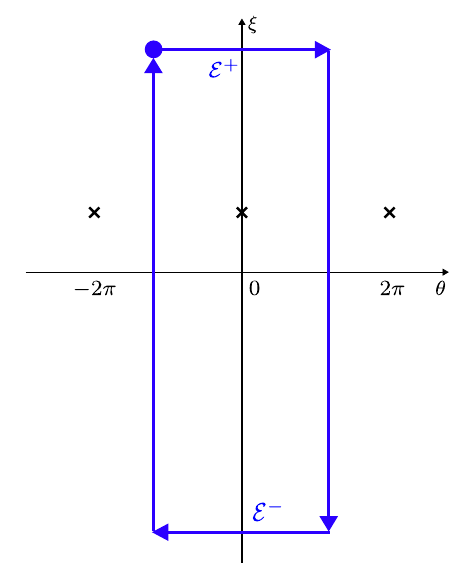}
		\caption{Clockwise loop}
		\label{clockwise}
	\end{subfigure}
	\begin{subfigure}[b]{0.45 \linewidth}
		\includegraphics[width=\linewidth]{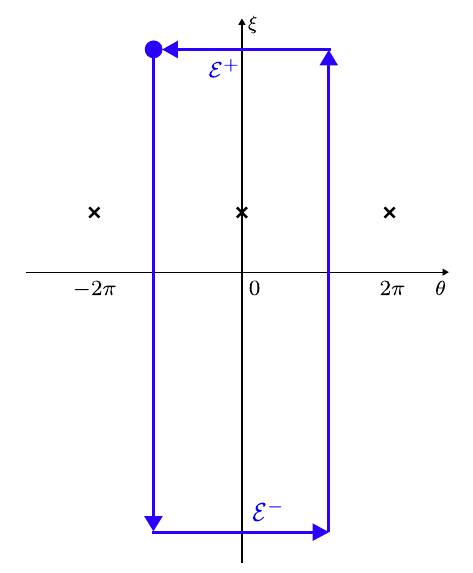}
		\caption{Anticlockwise loop}
		\label{anticlockwise}
	\end{subfigure}
	\centering
	\caption{Monodromy with paths around the singularity at $\theta = 0$ on the FI-theta space for $K_{Gr(2,4)}$. The crosses mark the singular point and the dot is the starting and end point of the paths.}
	\label{monodromyGr24}
\end{figure}

We start by considering a B-brane $\mathcal{B}\in\omega_{-2}$
where we choose $\omega_{-2}$ to be the sub-window of $\hat{\omega}_{-2}$,
after removing all instances of the modules $\mathcal{W}_{(2,1)}$ and $\mathcal{W}_{(3,2)}$ in every complex. The resulting sub-window is shown in Fig.
\ref{G24Win-2}, and we will start on the $\xi\gg 1$ phase (i.e. the basepoint
of our loop is located there). Therefore, this process involves two maps
between windows:
\begin{align}\label{Monproc24}
\omega_{-2}\xrightarrow{\mathcal{E}^{+}}\omega_{0}\xrightarrow{\mathcal{E}^{-}}
\omega_{-2},
\end{align}
where the objects $\mathcal{E}^{\pm}$, over the maps, denote the empty branes in
the phase $\pm\xi\gg 1$ used to map the objects between the windows at the two
ends of the arrow by taking cones. The path of the monodromy associated with \eqref{Monproc24} in the FI-theta space is depicted in Fig. \ref{clockwise}. In general, a single empty brane will not be enough, but in the example at hand, a single empty brane does the job.

The difference between our choices of $\omega_{-2}$ and $\omega_{0}$ are
the charges $(2,2)$ and $(-2,-2)$, so the empty brane $\mathcal{E}^{+}$
characterizing the first map in \eqref{Monproc24} can be chosen as:
\begin{align}
	\begin{array}{ccccccccccc}
		\mathcal{E}^+:~ \mathcal{W}_{(2,2)} & \rightarrow & \mathcal{W}_{(1,1)}^{\oplus 6} &
\rightarrow & \mathcal{W}_{(1,0)}^{\oplus 4} & \rightarrow & \mathcal{W}^{\oplus 4}_{(0,-1)} & \rightarrow &
\mathcal{W}_{(-1,-1)}^{\oplus 6} & \rightarrow & \mathcal{W}_{(-2,-2)}.
	\end{array}
	\label{PosEmpt24}
\end{align}

Upon mapping $\mathcal{B}$ from $\omega_{-2}$ to $\omega_{0}$, it can be
transported to the negative phase. Once in the negative phase, the brane needs
to be mapped to $\omega_{-2}$, using an empty brane in the $\xi\ll -1$ phase, as
the second map in \eqref{Monproc24} indicates. The brane $\mathcal{E}^{-}$ is
then given by
\begin{align}
	\begin{array}{ccc}
		\mathcal{E}^{-}:~ \mathcal{W}_{(2,2)} & \rightarrow &\mathcal{W}_{(-2,-2)}
	\end{array}.
	\label{NegEmpt24}
\end{align}

The journey back to the $\xi\gg 1$ phase can now be made through $\omega_{-2}$,
and $\mathcal{B}$ returns to $\omega_{-2}$ but altered by the journey into
$\mathcal{B}'$.

In the second map of \eqref{Monproc24}, the module $\mathcal{W}_{(-2,-2)}$ was swapped
out for $\mathcal{W}_{(2,2)}$. However, the only instances of $\mathcal{W}_{(-2,-2)}$ were those introduced by the cones with $\mathcal{E}^{+}$. Thus one can
combine $\mathcal{E}^{+}$ and $\mathcal{E}^{-}$ into a single brane to
produce the monodromy action of \eqref{Monproc24}:
\begin{align}
	\begin{array}{ccccccccccc}
		\mathcal{E}:~ \mathcal{W}_{(2,2)} & \rightarrow & \mathcal{W}_{(1,1)}^{\oplus 6} &
\rightarrow & \mathcal{W}_{(1,0)}^{\oplus 4} & \rightarrow & \mathcal{W}^{\oplus 4}_{(0,-1)} & \rightarrow &
\mathcal{W}_{(-1,-1)}^{\oplus 6} & \rightarrow & \mathcal{W}_{(2,2)}
	\end{array}.
	\label{MonActClock24}
\end{align}
In the positive phase we can identify the IR projection of
\eqref{MonActClock24} precisely with the object $\mathcal{E}$ in
\eqref{MonZCheck}:
\begin{align}
	\mathcal{E}\cong i_*(\mathrm{det}S)^{\otimes 2},\qquad \xi\gg 1.
\end{align}

We can check \eqref{Monproc24} more thoroughly by computing the hemisphere
partition function on the $\xi\gg 1$ phase. We do it for the generators of the
chosen basis given by \eqref{Rcomplex}, with
$\mu\in\{(0),(1),(1,1),(2),(2,1),(2,2)\}$. Denote $Z^{0,+}$ the zero instanton
sector of the hemisphere partition function in this basis. Then,
\begin{align}
	Z^{0,+} =
	\frac{16\pi^2}{3} \left( \begin{array}{c}
		t^3+12 i \pi  t^2-55 \pi ^2 t-92 i \pi ^3-66 \psi ^{(2)}(1) \\
		2 \left(t^3+9 i \pi  t^2-31 \pi ^2 t-39 i \pi ^3-66 \psi ^{(2)}(1)\right) \\
		t^3+6 i \pi  t^2-19 \pi ^2 t-22 i \pi ^3-66 \psi ^{(2)}(1) \\
		3\left(t^3+6 i \pi  t^2-11 \pi ^2 t-6 i \pi ^3-66 \psi ^{(2)}(1)\right)\\
		2 \left(t^3+3 i \pi  t^2-7 \pi ^2 t-5 i \pi ^3-66 \psi ^{(2)}(1)\right) \\
		t^3-7 \pi ^2 t-66 \psi ^{(2)}(1)
	\end{array} \right),
\end{align}
where $\psi ^{(2)}(z):=\frac{d^{3}}{z^{3}}\ln\Gamma(z)$, and upon monodromy action on the basis, denoted by $Z^{0,+}_M$, then
\begin{align}
	Z^{0,+}_M =
	\frac{16\pi^2}{3} \left( \begin{array}{c}
		-19 t^3+12 i \pi  t^2+85 \pi ^2 t-92 i \pi ^3+1254 \psi ^{(2)}(1) \\
		6 \left(-3 t^3+3 i \pi  t^2+13 \pi ^2 t-13 i \pi ^3+198 \psi ^{(2)}(1)\right) \\
		-5 t^3+6 i \pi  t^2+23 \pi ^2 t-22 i \pi ^3+330 \psi ^{(2)}(1) \\
		-7 t^3+18 i \pi  t^2+37 \pi ^2 t-18 i \pi ^3+462 \psi ^{(2)}(1) \\
		2 \left(-t^3+3 i \pi  t^2+7 \pi ^2 t-5 i \pi ^3+66 \psi ^{(2)}(1)\right) \\
		t^3-7 \pi ^2 t-66 \psi ^{(2)}(1)
	\end{array} \right).
\end{align}
Therefore, the expected monodromy transformation of the central charge in Eq. (\ref{MonZCheck}) is satisfied, with
\begin{align}
	\chi(\mathcal{B},\mathcal{E}) = (20,20,6,10,4,0).
	\label{G24ChiClock}
\end{align}
Then, the clockwise monodromy matrix in this basis takes the form
\begin{align}
	M_c = \left(
	\begin{array}{cccccc}
		1 & 0 & 0 & 0 & 0 & -20 \\
		0 & 1 & 0 & 0 & 0 & -20 \\
		0 & 0 & 1 & 0 & 0 & -6 \\
		0 & 0 & 0 & 1 & 0 & -10 \\
		0 & 0 & 0 & 0 & 1 & -4 \\
		0 & 0 & 0 & 0 & 0 & 1 \\
	\end{array}
	\right).
\end{align}

We can also perform the monodromy in the anticlockwise direction:
\begin{equation}\label{Monanti24}
\omega_{-2}\xrightarrow{\mathcal{E}^{-}}\omega_{0}\xrightarrow{\mathcal{E}^{+}
}
\omega_{-2}.
\end{equation}
The path of the monodromy associated with \eqref{Monanti24} in the FI-theta space is depicted in Fig. \ref{anticlockwise}. It can be shown that the monodromy matrix is given by
\begin{align}
	M_a = \left(
	\begin{array}{cccccc}
		1 & 0 & 0 & 0 & 0 & 20 \\
		0 & 1 & 0 & 0 & 0 & 20 \\
		0 & 0 & 1 & 0 & 0 & 6 \\
		0 & 0 & 0 & 1 & 0 & 10 \\
		0 & 0 & 0 & 0 & 1 & 4 \\
		0 & 0 & 0 & 0 & 0 & 1 \\
	\end{array}
	\right),
\end{align}
which is, as expected, the inverse of the monodromy matrix for the clockwise loop, $M_c = M_a^{-1}$. So, in summary, we were able to compute the monodromy, but we ended in a window with a different shape than we started. Naturally, we can use the branes (\ref{eeemptybrane}) to change the shape, but this is only possible in the $\xi\gg 1$ phase. The interesting remark is the following: once we choose a shape for the window $\omega_{s}$ for  $\xi\gg 1$, we can only loop clockwise or anti-clockwise depending on our choice, if we require that all the intermediate shapes are only of the types we declared valid, for $K_{Gr(2,4)}$. We revise this in more detail in the next subsection.

\subsection{A peculiarity with $K_{Gr(2,4)}$}\label{sec:peculiarity}

There are some loops that cannot be traversed for branes grade restricted to a particular valid window in the case of $K_{Gr(2,4)}$. In fact, upon choosing a valid subwindow $\omega_{s}\subset \hat{\omega}_{s}$, all grade-restricted branes cannot perform one out of the two possible loops that encircle a neighbouring singular point. In the previous subsection, the loop achievable by the brane $\mathcal{B}\in\omega_{-2}$ was presented, where it encircled the point at $\theta=0$. If it tried the other loop, encircling $\theta=-2\pi$, we would run into an invalid window as follows.

\begin{figure}[h]
	\begin{subfigure}[b]{0.45 \linewidth}
		\includegraphics[width=\linewidth]{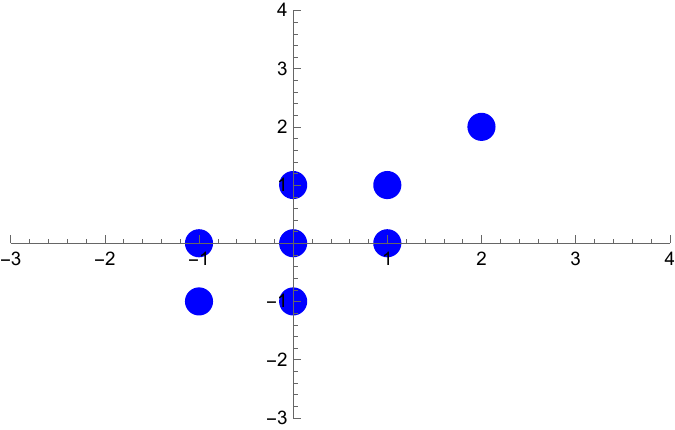}
		\caption{$\omega_{-2}$, type 1}
		\label{G24Win-2Type1}
	\end{subfigure}
	\begin{subfigure}[b]{0.45 \linewidth}
		\includegraphics[width=\linewidth]{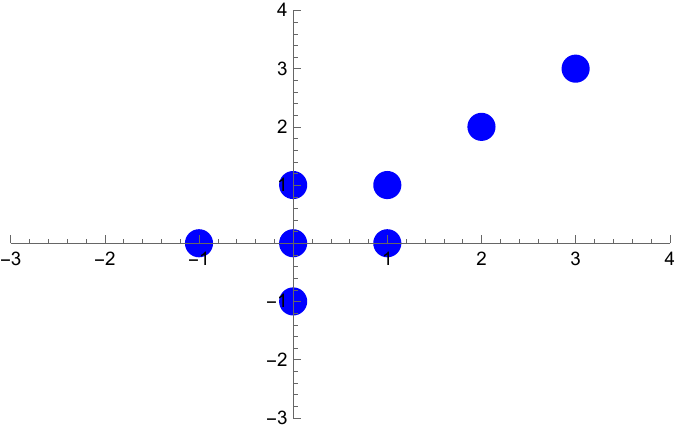}
		\caption{$\omega_{-4}$, invalid type.}
		\label{G24Win-4Invalid}
	\end{subfigure}
	\centering
	\caption{Attempting a monodromy loop around $\theta=-2\pi$, $K_{Gr(2,4)}$.}
	\label{G24InvalidMonodromy}
\end{figure}

Taking the brane $\mathcal{B}\in\omega_{-2}$ beginning in the positive phase, it can safely transport to the negative phase through the window $\omega_{-2}$. In order to loop around $\theta=-2\pi$, it would need to be re-grade restricted to satisfy the charge requirements of the neighbouring window $\omega_{-4}$, $-4\pi < \theta < -2\pi$. The only empty brane available for the job would be
\begin{align}
	\mathcal{W}_{(3,3)} \rightarrow \mathcal{W}_{(-1,-1)},
\end{align}
which would remove all instances of $\mathcal{W}_{(-1,-1)}$ and replace them with $\mathcal{W}_{(3,3)}$. This is the expected procedure -- the opposite shift of charges seen in the previous example. However, applying this to the brane $\mathcal{B}$ would leave it with the charge diagram in Fig. \ref{G24Win-4Invalid}. There is no empty brane in the negative phase that can alter the shape of this in such a way to bring it into a valid shape. Therefore, it cannot pass into the positive phase through $\omega_{-4}$, if we require every step to be through a valid window.

This is a unique restriction for the case of $N=4$, and does not occur for higher $N$. The reason this occurs is due to $K_{Gr(2,4)}$ having only one distinct singular point, where all other cases have at least two. Because of the absence of a singularity at $\theta=\pi$ in the $N=4$ case, as we shown in Sec. \ref{sec:evenresol}, we proposed discarding certain window shapes to avoid introducing an unphysical singularity: we defined 'invalid window shapes' in the $N=4$ case. Under the same reasoning, we do not find any need to discard particular window shapes in the $N>4$ case i.e. the cases of $N>4$ do not suffer this problem, and so grade-restricted branes in those models can be used to perform monodromies around either of the singularities bordering the window, for any choice of $\omega_{s}$. We remark that our analysis and proposal is based exclusively on the analysis of loops around straight paths that encircle all singularities above $\theta=0$ or $\theta=\pi$. It is possible that by analyzing curved paths in $N>4$, we are forced to also discard certain window shapes. However, for the families of monodromies analyzed in the present work, we do not find any contradiction.


\subsection{Example: $K_{Gr(2,5)}$}
\label{SecMonG25}

Eq. \eqref{singtheta} reveals two distinct singularities in $\mathcal{M}_{K}$,
for the case of $K_{Gr(2,5)}$: one located at $\theta=0$ and another at
$\theta=\pi$. Window shifting across $\theta=0$ requires the exchange
of an off-diagonal pair of charges, whilst the latter swaps between two points
lying on the diagonal. Each is investigated in turn.

In both cases, the brane $\mathcal{B}$
\begin{align}
	\begin{array}{c@{}c@{}c@{}c@{}c@{}c@{}c@{}c@{}c@{}c@{}c@{}c@{}c@{}c@{}c@{}}
		& & & & & & & & & & \mathcal{W}_{(3,3)} & & & & \\
		& & & & & & & & & & \oplus & & & & \\
		\mathcal{W}_{(3,3)} & \rightarrow & \mathcal{W}_{(3,2)}^{\oplus 5} & \rightarrow & \mathcal{W}_{(2,2)}^{\oplus 15} & \rightarrow & \mathcal{W}_{(2,1)}^{\oplus 40} & \rightarrow & \mathcal{W}_{(1,1)}^{\oplus 50} & \rightarrow & \mathcal{W}_{(-1,-1)}^{\oplus 15} & \rightarrow & \mathcal{W}_{(-1,-2)}^{\oplus 5} & \rightarrow & \mathcal{W}_{(0,-1)}^{\oplus 10} \\
		& & & & & & & & \oplus & & \oplus & & \oplus & & \\
		& & & & & & & & \mathcal{W}_{(2,1)}^{\oplus 50} & & \mathcal{W}_{(1,1)}^{\oplus 100} & & \mathcal{W}_{(0,0)}^{\oplus 50} & &
	\end{array}
	\label{25Mod33}
\end{align}
will be taken around a loop encircling a singular point. As it stands, it is grade restricted for the window $-2\pi < \theta < - \pi $, here denoted $\omega_{-2}$. Indeed \eqref{25Mod33} corresponds to the brane
\begin{equation}
	\mathcal{W}_{(3,3)} \rightarrow \mathcal{W}_{(-2,-2)}
\end{equation}
restricted to $\omega_{-2}$ by taking successive cones. It will serve as the initial state for all journeys in this subsection.


\subsubsection{$\theta=0$ - the off-diagonal exchange}

\begin{figure}[h]
	\begin{subfigure}[b]{0.45 \linewidth}
		\includegraphics[width=\linewidth]{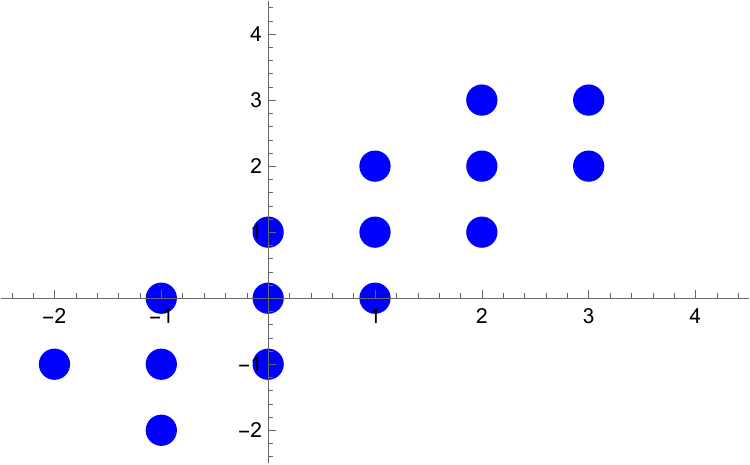}
		\caption{$\omega_{-2}$}
	\end{subfigure}
	\begin{subfigure}[b]{0.45 \linewidth}
		\includegraphics[width=\linewidth]{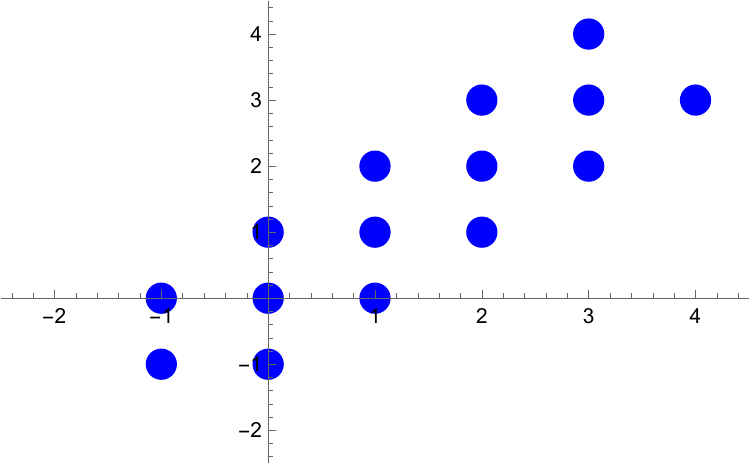}
		\caption{$\omega_{-3}$}
	\end{subfigure}
	\centering
	\caption{The charge restrictions for two windows of $K_{Gr(2,5)}$.}
	\label{G25MonWins}
\end{figure}

The first example sees the brane $\mathcal{B}$ making a round trip across the moduli space by setting off from the deep positive phase into the negative phase through $\omega_{-2}$ by virtue of it already being grade restricted for that window. Thus it travels unhindered and, after its arrival in the negative phase, it seeks return passage through $\omega_{-3}$. To do so, it must be re-grade restricted in order to satisfy that window's separate charge conditions. A simple change is needed: the charges $(-2,-1)$ and $(-1,-2)$ must be exchanged for $(3,4)$ and $(4,3)$. This is equivalent to replacing the module $\mathcal{W}_{(-1,-2)}$ with $\mathcal{W}_{(4,3)}$.
The charge restrictions for each of these windows are shown in Fig. \ref{G25MonWins}.


The task is done by binding to $\mathcal{B}$ the following empty brane in the negative phase,
\begin{align}
	\begin{array}{ccc}
		\mathcal{E}^-_0: \mathcal{W}_{(4,3)} & \overset{p} \rightarrow & \mathcal{W}_{(-1,-2)}
		\label{NegEmpt25}
	\end{array}.
\end{align}
By binding $\mathcal{E}^-_0$ to $\mathcal{B}$ at every instance of the module $\mathcal{W}_{(-1,-2)}$, brane-antibrane annihilation removes it completely, and so the necessary exchange is achieved. The resulting brane is then
\begin{align}
	\begin{array}{c@{}c@{}c@{}c@{}c@{}c@{}c@{}c@{}c@{}c@{}c@{}c@{}c@{}c@{}c@{}}
		& & & & & & & & & & \mathcal{W}_{(3,3)} & & \mathcal{W}_{(4,3)}^{\oplus 5} & \rightarrow & \cancel{\mathcal{W}_{(-1,-2)}^{\oplus 5}} \\
		& & & & & & & & & & \oplus & & \oplus & \nearrow_{\text{Id}} & \oplus \\
		\mathcal{W}_{(3,3)} & \rightarrow & \mathcal{W}_{(3,2)}^{\oplus 5} & \rightarrow & \mathcal{W}_{(2,2)}^{\oplus 15} & \rightarrow & \mathcal{W}_{(2,1)}^{\oplus 40} & \rightarrow & \mathcal{W}_{(1,1)}^{\oplus 50} & \rightarrow & \mathcal{W}_{(-1,-1)}^{\oplus 15} & \rightarrow & \cancel{\mathcal{W}_{(-1,-2)}^{\oplus 5}} & \rightarrow & \mathcal{W}_{(0,-1)}^{\oplus 10} \\
		& & & & & & & & \oplus & & \oplus & & \oplus & & \\
		& & & & & & & & \mathcal{W}_{(2,1)}^{\oplus 50} & & \mathcal{W}_{(1,1)}^{\oplus 100} & & \mathcal{W}_{(0,0)}^{\oplus 50} & & 	
	\end{array}
\end{align}
and is denoted by $\mathcal{B}'$. It is now grade restricted for the window $\omega_{-3}$ and ready to return to the positive phase.

Once its return leg is complete, the brane, now $\mathcal{B}'$ and back in the positive phase, must be re-grade restricted once more to return to its original state of satisfying the charge restriction of $\omega_{-2}$ before it can complete its journey. This amounts to reversing the change to its charges made above whilst $\mathcal{B}'$ was in the negative phase. However, it is now in the positive phase and thus needs to be treated by a brane that is empty in the positive phase. It turns out the empty brane
\begin{align}
	\begin{array}{c@{}c@{}c@{}c@{}c@{}c@{}c@{}c@{}c@{}c@{}c@{}c@{}c@{}c@{}c@{}}
		\mathcal{E}^+_0:~ \mathcal{W}_{(4,3)} & \rightarrow & \mathcal{W}_{(3,3)}^{\oplus 5} & \rightarrow & \mathcal{W}_{(2,2)}^{\oplus 10} & \rightarrow & \mathcal{W}_{(2,1)}^{\oplus 5} & \rightarrow & \mathcal{W}_{(1,0)}^{\oplus 5} & \rightarrow & \mathcal{W}_{(0,0)}^{\oplus 10} & \rightarrow & \mathcal{W}_{(-1,-1)}^{\oplus 5} & \rightarrow & \mathcal{W}_{(-1,-2)}
		\label{PosEmpt25}
	\end{array}
\end{align}
will do the job. 
The resulting brane, which is now denoted $\mathcal{B}''$, is
\begin{align}
	\begin{array}{c@{}c@{}c@{}c@{}c@{}c@{}c@{}c@{}c@{}c@{}c@{}c@{}c@{}}
		& & & & & & & & \bcancel{\mathcal{W}_{(4,3)}^{\oplus 5}} & \rightarrow & \mathcal{W}_{(3,3)}^{\oplus 25} & \rightarrow & ... \\
		& & & & & & & \nearrow_{\text{Id}} & \oplus & & & &  \\
		& & & & \mathcal{W}_{(3,3)} & & \bcancel{\mathcal{W}_{(4,3)}^{\oplus 5}} & \rightarrow & \cancel{\mathcal{W}_{(-1,-2)}^{\oplus 5}} \\
		& & & & \oplus & & \oplus & \nearrow_{\text{Id}} & \oplus & & & & \\
		... & \rightarrow & \mathcal{W}_{(1,1)}^{\oplus 50} & \rightarrow & \mathcal{W}_{(-1,-1)}^{\oplus 15} & \rightarrow & \cancel{\mathcal{W}_{(-1,-2)}^{\oplus 5}} & \rightarrow & \mathcal{W}_{(0,-1)}^{\oplus 10} & & & & \\
		& & \oplus & & \oplus & & \oplus & & & & & & \\
		& & \mathcal{W}_{(2,1)}^{\oplus 50} & & \mathcal{W}_{(1,1)}^{\oplus 100} & & \mathcal{W}_{(0,0)}^{\oplus 50} & & & & & &
	\end{array},
\end{align}
where omissions have been made to fit the more pertinent parts of the brane. The ellipsis on the left-hand side represents the left tail of the brane $\mathcal{B}'$, while the ellipsis of the right represents the right tail of five copies of the empty brane $\mathcal{E}^+_0$.

The brane $\mathcal{B}$ has now completed its journey, and returns as $\mathcal{B}''$ which, after all the brane-antibrane annihilation occurs, reduces to
\begin{align}
	\begin{array}{c@{}c@{}c@{}c@{}c@{}c@{}c@{}c@{}c@{}c@{}c@{}c@{}c@{}c@{}c@{}}
		& & & & \mathbfcal{W}_{(3,3)} & & & & & & \\
		& & & & \oplus & & & & & & & & & & \\
		... & \rightarrow & \mathbfcal{W}_{(1,1)}^{\oplus 50} & \rightarrow & \mathbfcal{W}_{(-1,-1)}^{\oplus 15} & \rightarrow & \mathbfcal{W}_{(0,0)}^{\oplus 50} & \rightarrow & \mathbfcal{W}_{(0,-1)}^{\oplus 10} & \rightarrow & \mathbfcal{W}_{(3,3)}^{\oplus 25} & \rightarrow & \mathcal{W}_{(2,2)}^{\oplus 50} & \rightarrow & ...\\
		& & \oplus & & \oplus & & & & & & & & & & \\
		& & \mathbfcal{W}_{(2,1)}^{\oplus 50} & & \mathbfcal{W}_{(1,1)}^{\oplus 100} & & & & & & & & & &
	\end{array}.
\end{align}
Here the boldface highlights the modules that were members of the original brane $\mathcal{B}$. The overall change is then clear: the entire monodromy action can be summed up by the binding of a single brane:
\begin{align}
	\mathcal{E}_{0}:=\begin{array}{c@{}c@{}c@{}c@{}c@{}c@{}c@{}c@{}c@{}c@{}c@{}c@{}c@{}c@{}c@{}}
		\underline{\mathcal{W}_{(-1,-2)}} & \rightarrow & \mathcal{W}_{(3,3)}^{\oplus 5} & \rightarrow & \mathcal{W}_{(2,2)}^{\oplus 10} & \rightarrow & \mathcal{W}_{(2,1)}^{\oplus 5} & \rightarrow & \mathcal{W}_{(1,0)}^{\oplus 5} & \rightarrow & \mathcal{W}_{(0,0)}^{\oplus 10} & \rightarrow & \mathcal{W}_{(-1,-1)}^{\oplus 5} & \rightarrow & \mathcal{W}_{(-1,-2)}
	\end{array}.
\end{align}
The underlined module is the one that is bound to the brane undergoing the monodromy journey, targeting like modules in the travelling brane.

The general monodromy action is one where a brane, grade restricted for the window $\omega_{-2}$, is repeatedly bound with the brane $\mathcal{E}_{0}$ at every instance of the module $\mathcal{W}_{(-1,-2)}$.

As in the $N=4$ case, the full set of R-basis branes, once appropriately grade restricted, can undergo this journey around the moduli space. With the basis of representations $\{(0,0), (1,0), (1,1),$ $(2,0), (2,1), (2,2), (3,0), (3,1), (3,2), (3,3)\}$, their central charges (computed by the hemisphere partition function) $Z^{+}$ at the beginning of the trip and the central charges $Z^{+}_T$ at the end can be compared. This is a residue calculation with an infinite number of poles, whose terms are graded by powers of $e^{-t}$, but the first terms in the expansion will suffice in making the comparison.

The result shows that the expected monodromy transformation of the central charge in Eq.  (\ref{MonZCheck}) is satisfied, with
\begin{align}
	\chi(\mathcal{B}, \mathcal{E}_{0}) = -(560,595,175,400,155,40,180,75,25,5).
	\label{G25ChiClock1}
\end{align}
The monodromy matrix of the central charges satisfying $Z^{+}_T = {M_{0,c}} \cdot Z^{+}$ is then
\begin{align}
	M_{0,c} = \left(
	\begin{array}{cccccccccc}
		1 & 0 & 0 & 560 & -2800 & 5600 & 0 & 0 & 0 & -2800 \\
		0 & 1 & 0 & 595 & -2975 & 5950 & 0 & 0 & 0 & -2975 \\
		0 & 0 & 1 & 175 & -875 & 1750 & 0 & 0 & 0 & -875 \\
		0 & 0 & 0 & 401 & -2000 & 4000 & 0 & 0 & 0 & -2000 \\
		0 & 0 & 0 & 155 & -774 & 1550 & 0 & 0 & 0 & -775 \\
		0 & 0 & 0 & 40 & -200 & 401 & 0 & 0 & 0 & -200 \\
		0 & 0 & 0 & 180 & -900 & 1800 & 1 & 0 & 0 & -900 \\
		0 & 0 & 0 & 75 & -375 & 750 & 0 & 1 & 0 & -375 \\
		0 & 0 & 0 & 25 & -125 & 250 & 0 & 0 & 1 & -125 \\
		0 & 0 & 0 & 5 & -25 & 50 & 0 & 0 & 0 & -24 \\
	\end{array}
	\right).
\end{align}
It can be shown that if we reverse the direction, making an anticlockwise loop through the same pair of windows $\omega_{-3}$ and $\omega_{-2}$, then the resulting monodromy matrix $M_{0,a}$ is indeed the inverse of $M_{0,c}$.

\subsubsection{$\theta=\pi$ - the diagonal exchange}

\begin{figure}[h]
	\begin{subfigure}[b]{0.45 \linewidth}
		\includegraphics[width=\linewidth]{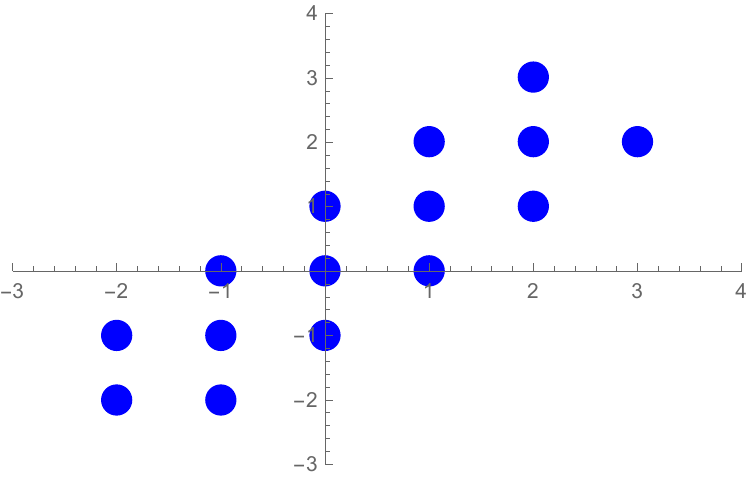}
		\caption{$\omega_{-1}$}
	\end{subfigure}
	\begin{subfigure}[b]{0.45 \linewidth}
		\includegraphics[width=\linewidth]{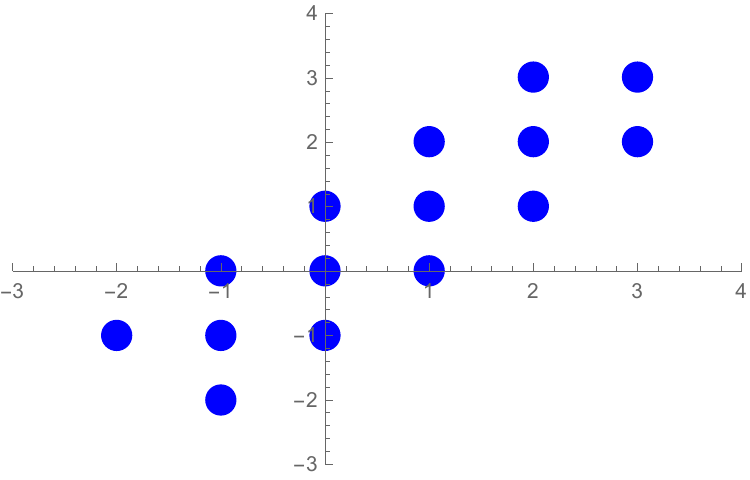}
		\caption{$\omega_{-2}$}
	\end{subfigure}
	\centering
	\caption{The charge restrictions for two windows of $K_{Gr(2,5)}$.}
	\label{G25MonWins2}
\end{figure}

There is another singular point: one located at $\theta = \pi$. The process of encompassing this singularity is very similar to the previous one, but with different empty branes targeting different modules. The windows are naturally different too: to encircle the singular point, a brane would travel through $\omega_{-1}$ and $\omega_{-2}$, the former of which denotes $-\pi < \theta < 0$. The charge restrictions for these windows are shown in Fig. \ref{G25MonWins2}.

The brane $\mathcal{B}$ from the previous section can travel this loop also, but the process results in branes too bulky to print here. The crucial aspects are printed here, however, which are the empty branes and how they are combined into a single, overall monodromy action.

The journey begins in the deep positive phase and leads $\mathcal{B}$ through $\omega_{-1}$ and into the deep negative phase. It returns to the positive phase through $\omega_{-2}$, thus looping around the singular point located at $\theta = -\pi$, which is equivalent to the singular point at $\theta = \pi$ by periodicity.

The brane is originally grade restricted for $\omega_{-2}$, and so must first be re-grade restricted in the positive phase, exchanging $\mathcal{W}_{(3,3)}$ for $\mathcal{W}_{(-2,-2)}$ using the empty brane $\mathcal{E}^+_\pi$:
\begin{align}
	\begin{array}{c@{}c@{}c@{}c@{}c@{}c@{}c@{}c@{}c@{}c@{}c@{}c@{}c@{}c@{}c@{}}
		\mathcal{W}_{(3,3)} & \rightarrow & \mathcal{W}_{(3,2)}^{\oplus 5} & \rightarrow & \mathcal{W}_{(2,2) \oplus 15} & \rightarrow & \mathcal{W}_{(2,1) \oplus 40} & \rightarrow & \mathcal{W}_{(1,1) \oplus 50} & \rightarrow & \mathcal{W}_{(-1,-1) \oplus 15} & \rightarrow & \mathcal{W}_{(-1,-2)}^{\oplus 5} & \rightarrow & \mathcal{W}_{(-2,-2)} \\
		& & & & \oplus & & \oplus & & \oplus & & \oplus & & & & \\
		& & & & \mathcal{W}_{(2,1)}^{\oplus 50} & & \mathcal{W}_{(1,1)}^{\oplus 100} & & \mathcal{W}_{(0,0)}^{\oplus 50} & & \mathcal{W}_{(0,-1)}^{\oplus 10} & & & &
	\end{array}.
\end{align}
This is bound at all instances of the module $\mathcal{W}_{(3,3)}$. Once done, the brane is now free to travel through $\omega_{-1}$ and towards the deep negative phase.

Now the brane prepares for its return trip through $\omega_{-2}$, and so must be re-grade restricted once more to undo the exchange that took place above. To do this in the negative phase, the following empty brane is bound,
\begin{align}
	\begin{array}{ccc}
		\mathcal{E}^-_\pi:~ \mathcal{W}_{(3,3)} & \rightarrow & \mathcal{W}_{(-2,-2)}
	\end{array}.
\end{align}
Once this brane is bound to all instances of the introduced module $\mathcal{W}_{(-2,-2)}$, the composite is free to return to its initial state in the positive phase.

The only appearance of $\mathcal{W}_{(-2,-2)}$ is due to the presence of the bound branes $\mathcal{E}^+_\pi$, so the whole procedure can be described with the application of a single composite brane $\mathcal{E}_{\pi}$:
\begin{align}
	\mathcal{E}_{\pi}=\begin{array}{c@{}c@{}c@{}c@{}c@{}c@{}c@{}c@{}c@{}c@{}c@{}c@{}c@{}c@{}c@{}}
		\underline{\mathcal{W}_{(3,3)}} & \rightarrow & \mathcal{W}_{(3,2)}^{\oplus 5} & \rightarrow & \mathcal{W}_{(2,2)}^{\oplus 15} & \rightarrow & \mathcal{W}_{(2,1)}^{\oplus 40} & \rightarrow & \mathcal{W}_{(1,1)}^{\oplus 50} & \rightarrow & \mathcal{W}_{(-1,-1)}^{\oplus 15} & \rightarrow & \mathcal{W}_{(-1,-2)}^{\oplus 5} & \rightarrow & \mathcal{W}_{(3,3)} \\
		& & & & \oplus & & \oplus & & \oplus & & \oplus & & & & \\
		& & & & \mathcal{W}_{(2,1)}^{\oplus 50} & & \mathcal{W}_{(1,1)}^{\oplus 100} & & \mathcal{W}_{(0,0)}^{\oplus 50} & & \mathcal{W}_{(0,-1)}^{\oplus 10} & & & &
	\end{array},
\end{align}
where the underlined module is the point of contact when binding to a brane.

The central charges of the R-basis branes can be seen to change as they are taken around this loop. Using the same basis as in the previous case,
the expected monodromy matrix of the central charge in Eq. (\ref{MonZCheck}) is satisfied, with
\begin{align}
	\chi(\mathcal{B},\mathcal{E}_{\pi}) = (175,175,50,105,40,10,35,15,5,0).
	\label{G25ChiClock2}
\end{align}
In addition, one can calculate the monodromy matrix to be
\begin{align}
	M_{\pi,c} = \left(
	\begin{array}{cccccccccc}
		1 & 0 & 0 & 0 & 0 & 0 & 0 & 0 & 0 & 175 \\
		0 & 1 & 0 & 0 & 0 & 0 & 0 & 0 & 0 & 175 \\
		0 & 0 & 1 & 0 & 0 & 0 & 0 & 0 & 0 & 50 \\
		0 & 0 & 0 & 1 & 0 & 0 & 0 & 0 & 0 & 105 \\
		0 & 0 & 0 & 0 & 1 & 0 & 0 & 0 & 0 & 40 \\
		0 & 0 & 0 & 0 & 0 & 1 & 0 & 0 & 0 & 10 \\
		0 & 0 & 0 & 0 & 0 & 0 & 1 & 0 & 0 & 35 \\
		0 & 0 & 0 & 0 & 0 & 0 & 0 & 1 & 0 & 15 \\
		0 & 0 & 0 & 0 & 0 & 0 & 0 & 0 & 1 & 5 \\
		0 & 0 & 0 & 0 & 0 & 0 & 0 & 0 & 0 & 1 \\
	\end{array}
	\right).
\end{align}
Again, in the reverse direction, one can check that the monodromy matrix is the inverse of $M_{\pi,c}$.

\subsection{General monodromy actions}\label{sec:genmonaction}

In this section we will outline some general aspects about monodromies for general $N$. We will limit ourselves to present some general properties that we believe can help building an understanding of the properties shared between all possible values of $N$, making the distinction between the odd and even cases.

We note that every window can be understood as a collection of `diagonals' that come in pairs (because of Weyl invariance of the windows). That means charges on a line of slope $1$, shifted $n$ steps from the diagonal. More precisely the diagonal corresponds to charges satisfying $q^1=q^{2}$, and $n=0$ and the charges belonging to a pair of `off-diagonals' labelled by $n\in \mathbb{N}$ will satisfy $q^2=q^{1}\pm n$. We will refer to these lines as diagonals with $n\in \mathbb{N}$ shifts.

Then, for $N$ odd, in any window category each diagonal, with any number of shifts $n$, contains $N$ charges. Moreover $n\leq N-1$. For $N$ even, any admissible sub-window categories will have diagonals of length $N$, except for the outermost one, i.e. the one shifted $n=N-2$ times, which contains $\frac{N}{2}$ charges.

When we perform the monodromy around a singularity, depending on the location of it, namely $\theta=0$ or $\theta=\pi$, the adjacent windows will differ by their charges at $n$-shifted diagonals, following the pattern:

For odd $N$, when crossing over the singularity at
\begin{itemize}
	\item $\theta = 0$, charges on diagonals with odd $n$ shifts differ;
	\item $\theta = \pi$, charges on diagonals with even $n$ shifts differ.
\end{itemize}
The situation is opposite for even $N$. In this case, when crossing over the singularity at
\begin{itemize}
	\item $\theta = 0$, charges on diagonals with even $n$ shifts differ;
	\item $\theta = \pi$, charges on diagonals with odd $n$ shifts differ.
\end{itemize}

\begin{figure}[h]
	\begin{subfigure}[b]{0.32 \linewidth}
		\includegraphics[width=\linewidth]{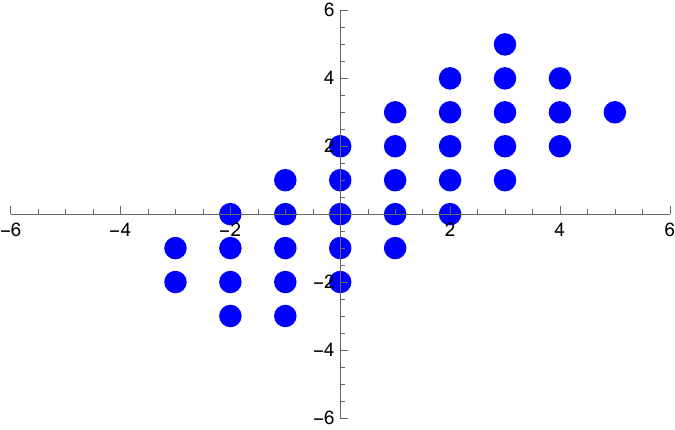}
		\caption{$\omega_{-2}$}
	\end{subfigure}
	\begin{subfigure}[b]{0.32 \linewidth}
		\includegraphics[width=\linewidth]{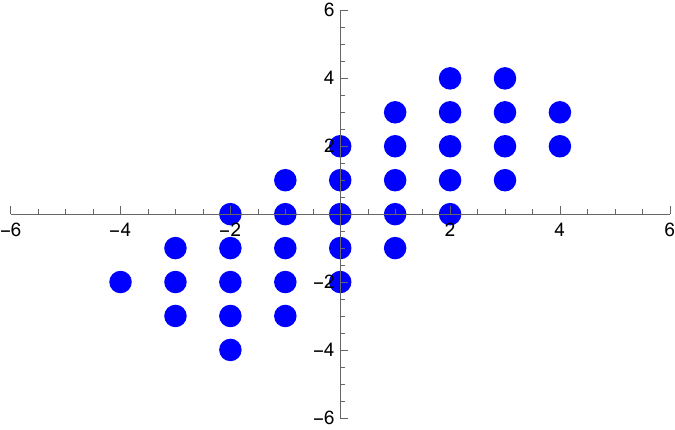}
		\caption{$\omega_{-1}$}
	\end{subfigure}
	\begin{subfigure}[b]{0.32 \linewidth}
		\includegraphics[width=\linewidth]{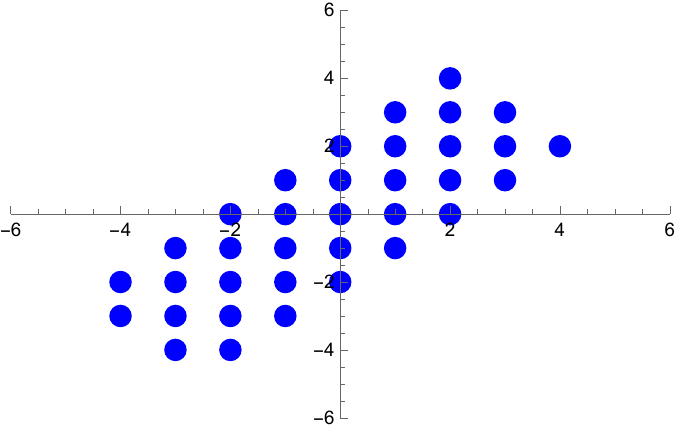}
		\caption{$\omega_{-0}$}
	\end{subfigure}
	\centering
	\caption{The charge restrictions for three windows of $K_{Gr(2,7)}$.}
	\label{G27Wins1}
\end{figure}

As an example, we show a sample in Fig. \ref{G27Wins1} of three adjacent windows for the case $N=7$, namely $\omega_{-2}$, $\omega_{-1}$ and $\omega_{0}$. Each diagonal has length $7$, and the charge diagram has three shifted diagonals with $n=0,1,2$. For example, a B-brane grade restricted to $\omega_{-1}$ when undergoing monodromy around the singularity at $\theta=0$ must be re-graded to fit $\omega_{0}$ by swapping a charge on the $n=1$ shifted diagonals: the module $\mathcal{W}_{(4,3)}$ is replaced with $\mathcal{W}_{(-3,-4)}$.

Traversing the other singularity at $\theta=-\pi$ instead requires re-grading to the window $\omega_{-2}$, which differs from $\omega_{-1}$ by the charges on the $n=0$ and $n=1$ shifted diagonals. This requires two modules to be replaced: $\mathcal{W}_{(-3,-3)}$ should be replaced with $\mathcal{W}_{(4,4)}$, and $\mathcal{W}_{(-2,-4)}$ should be replaced with $\mathcal{W}_{(5,3)}$.

\subsubsection{Odd $N$}
\label{OddNBraneGen}
For odd $N$, the empty branes that are used for window shifting are as follows. We denote $\mathcal{E}^\pm$ for an empty brane on the positive/negative phase.

For shifting between windows in the $\xi\gg 1$ phase, it suffices to use empty branes constructed from exact sequences of the form $L_{\mu}E$ (and their twists by $\mathrm{det}S$), where $E:= S\rightarrow V\rightarrow Q$ denotes the Euler sequence. They can be constructed algorithmically \cite{thesislucy} for arbitrary $N$ odd, but here we will just argue this holds by presenting an example in detail in Sec. \ref{Examples25}.

In the $\xi\ll -1$ phase, the empty branes needed for shifting between windows are quite simple, specifically, we need to take cones by any subset of the following (and their twists by $\mathcal{W}_{(1,1)}$):
\begin{align}\label{emptyBoddd}
	\mathcal{E}^-_{n} = \mathcal{W}_{(a,a-n)} \rightarrow \mathcal{W}_{(a-N,a-n-N)},\qquad a:=\frac{N-1}{2},n=0,\ldots,a-1.
\end{align}
The empty branes $\mathcal{E}^{-}_{n}$ can be twisted such that most of its charges fit inside a window and the only charges outside, belong to the $n$-shifted diagonal (using the terminology of Sec. \ref{sec:genmonaction}).




\subsubsection{Even $N$}
\label{EvenNBraneGen}
The even $N$ case presents an additional complexity due to each window possessing multiple valid charge configurations as discussed in Sec. \ref{sec:evenresol}. For instance, Fig. \ref{G24MonWins2} presented the two possible shapes of the charge diagram for $N=4$ and a selection of those for $N=6$ is shown in Fig. \ref{G26PossWins}. For $N=4$ we encounter the peculiarity that, depending on whether we want to perform the monodromy clockwise or counterclockwise, we need to choose different shapes of the windows. This is described in detail in Sec. \ref{G24Mon}. This direction-depending choice was necessary in order to avoid passing through invalid window configurations, as we declared them in Sec. \ref{sec:evenresol}.

In the $N$ even, $N>4$ cases, we do not encounter the same issue as of $N=4$. In this case any of the valid window configurations contained in $\hat{\omega}_{s}$ can be chosen and we can perform the monodromy, clockwise or counterclockwise,  without going through invalid window configurations.


For the negative phase, shifting between windows requires taking cones by any subset of the following (and their twists by $\mathcal{W}_{(1,1)}$):
\begin{align}
	\mathcal{E}^-_n = \mathcal{W}_{(a,a-n)} \rightarrow \mathcal{W}_{(a-N,a-n-N)}, \qquad a:=\frac{N}{2}-1,n=0,\ldots,a-1,
\end{align}
the subindex $n$ has the same meaning as in Sec. \ref{OddNBraneGen}. The empty branes $\mathcal{E}^-_n$ will always map valid windows into valid windows for $N>4$ and in the $\xi\gg 1$ phase, we always have enough empty branes to change from one valid window to another. As in the even $N$ case we only argue for this by showing an example in detail in Sec. \ref{subsecG26}, but a more detailed algorithm can be found in \cite{thesislucy}.


\section{The $\xi\ll -1$ phase for $N=4$}\label{sec:negphase}

As discussed in Sec. \ref{sec:setup}, the $K_{Gr(k,N)}$ GLSM under discussion with matter content \eqref{mattercontent} RG flows to a NLSM with target space $K_{Gr(k,N)}$ in the $\xi\gg 1$ phase. However, an explicit IR description in the $\xi\ll-1$ phase is still missing. The classical solution to the D-term equations is given by a $\mathbb{Z}_N$-orbifold of the affine cone of the Grassmannian $Gr(k,N)$ under the Pl\"ucker embedding. Let us denote the affine cone by $CGr(k,N)$.

In this section, we focus only on the $k=2$, $N=4$ case. In the $\xi\ll -1$ phase of the $U(2)$ GLSM describing $K_{Gr(2,4)}$, the $P$ field acquires a nonzero vev, which breaks $U(2)$ to the subgroup $H$ given by
\[
H = \left\{ g\in U(2):(\det g)^{4}=1\right\},
\]
where the $SU(2)$ fundamentals $\Phi^{\alpha}_i$ all have weight one under the quotient group $\mathbb{Z}_4$. The gauge invariant baryons describe the classical solution to the D-term equations:
\[
B_{ij} = \epsilon_{\alpha\beta} \Phi^{\alpha}_i \Phi^{\beta}_j, \quad 1 \leq i<j \leq 4,
\]
which all transform in the $\mathbb{Z}_4$ representation of weight two. As we reviewed in Sec. \ref{sec:regularity} there is also a non-compact Coulomb branch in this phase, as shown in \cite{Hori:2006dk}. However in this particular model, we can write a dual abelian GLSM with gauge group $U(1)$ modelling the line bundle $K_{Gr(2,4)}$. This model does not have a non-compact Coulomb branch in its negative phase, and its geometric phase corresponds to the NLSM with target space $K_{Gr(2,4)}$ embedded in $\mathcal{O}_{\mathbb{P}^{5}}(-2)\oplus \mathcal{O}_{\mathbb{P}^{5}}(-4)$. Hence we propose, that at least in the case  $k=2$, $N=4$, the $\xi\ll -1$ phase of the $U(2)$ GLSM must have a pure Higgs branch unaffected by the presence of the non-compact Coulomb branch and whose category of B-branes is equivalent to $D(K_{Gr(2,4)})$.

The dual abelian description\footnote{The similar $U(1)$ model describing a Calabi-Yau hypersurface in $Gr(2,4)$ was discussed in Sec. 4.6 of \cite{Hori:2006dk}.} is given by a $U((1)$ gauge theory with six chiral fields $X_{ij} = -X_{ji}, i, j=1,2,3,4,$ of charge 1 and two chiral fields $P_1$ and $P_2$ of charge $-2$ and $-4$ respectively. There is a superpotential $W=P_1 G(X)$, where
\begin{equation}
G(X) = X_{12} X_{34}-X_{13} X_{24}+X_{14} X_{23}.
\end{equation}
Let $r$ be the FI-parameter of this $U(1)$ theory. Then at $r \gg 1$, the low-energy theory has target space $\mathrm{Tot}(\mathcal{O}_{\mathcal{X}}(-4) \rightarrow \mathcal{X}) \cong K_{Gr(2,4)}$, where $\mathcal{X}$ is the quadric defined by $\mathcal{X} = \{ X \in \mathbb{P}^5 ~|~ G(X)=0 \}$, which is the image of the Grassmannian $Gr(2,4)$ under the Pl\"ucker embedding, and $\mathcal{O}_{\mathcal{X}}(-4)$ denotes the line bundle $\mathcal{O}_{\mathbb{P}^5}(-4)$ pulled back to $\mathcal{X}$ via the embedding. At $r \ll -1$, $P_1$ and $P_2$ cannot vanish simultaneously, and the low-energy theory can be described by a hybrid model on a rank-6 vector bundle over the weighted projective space $\mathbb{WP}^{1,2}$, with $X_{ij}$ and $(P_1, P_2)$ being the fibre and base coordinates respectively. At each point $(P_1, P_2)$ on the weighted projective space, there is a $\mathbb{Z}_2$-orbifold of the LG model defined on the fibre with superpotential $W=P_1 G(X)$. Since $G$ is quadratic, the fibre fields are massive except at at the point $P_1=0$. This is a sign of singularity in the $\xi \rightarrow -\infty$ limit as discussed in \cite{Hori:2006dk}.

We can compare the window category of the dual abelian theory with the window category we obtained in Sec. \ref{sec:evenresol} for $K_{Gr(2,4)}$. Following \cite{Herbst:2008jq}, with an appropriate choice of the theta-angle, the grade restriction rule of the dual $U(1)$ theory selects gauge charges $q$ satisfying $-2 \leq q \leq 3$. Let $\mathcal{M}_{(q)}$ be the one-dimensional $\mathbb{C}[X_{12},\cdots, X_{34}, P_1, P_2]$-module with $U(1)$ charge $q$, a set of generators of the window category can then be chosen to be $\{\mathcal{A}_0, \mathcal{A}_1,\mathcal{A}_2, \mathcal{A}_3,\mathcal{F}_{+}, \mathcal{F}_{-}\}$, where $\mathcal{A}_m$ is the matrix factorization
\begin{equation}
\xymatrix{
{\mathcal{M}_{(m-2)}} \ar@<0.5ex>[rr]^{G(X)} & & \mathcal{M}_{(m)} \ar@<0.5ex>[ll]^{P_1}},\quad m=0,1,2,3.
\end{equation}
$\mathcal{F}_{+}$ and $\mathcal{F}_{-}$ are the matrix factorizations
\begin{equation}\label{fplus}
\xymatrix{
{\mathcal{M}_{(-1)}^{\oplus 4}} \ar@<0.5ex>[rr]^{\psi_+(X)} & & \mathcal{M}_{(0)}^{\oplus 4} \ar@<0.5ex>[ll]^{P_1 \cdot \psi_-(X)}}
\end{equation}
and
\begin{equation}\label{fminus}
\xymatrix{
{\mathcal{M}_{(0)}^{\oplus 4}} \ar@<0.5ex>[rr]^{P_1 \cdot \psi_+(X)} & & \mathcal{M}_{(-1)}^{\oplus 4} \ar@<0.5ex>[ll]^{\psi_-(X)}}
\end{equation}
respectively, where $\psi_{\pm}$ are the linear maps such that $\psi_+ \circ \psi_- = \psi_- \circ \psi_+ = G(X) \cdot \mathrm{id}$. An explicit matrix form of $\psi_{\pm}$ is given as follows:
\[
\psi_+(X) = \left(
\begin{array}{cccc}
X_{34} &  -X_{24} & X_{23} & 0 \\
-X_{13} & X_{12} & 0 & X_{23} \\
-X_{14} & 0 & X_{12} & X_{24} \\
0 & -X_{14} & X_{13} & X_{34}
\end{array}\right), \quad \quad
\psi_-(X) = \left(
\begin{array}{cccc}
X_{12} & X_{24} & -X_{23} & 0 \\
X_{13} & X_{34} & 0 & -X_{23} \\
X_{14} & 0 & X_{34} & -X_{24} \\
0 & X_{14} & -X_{13} & X_{12}
\end{array}\right).
\]
At $r \gg 1$, $P_1 \rightarrow 0$ and $\mathcal{M}_{(q)} \rightarrow \mathcal{O}_{\mathbb{P}^5}(q)$ at low energy. From the resolutions in $D(\mathbb{P}^5)$
\[
\mathcal{O}_{\mathbb{P}^5}(m-2) \stackrel{G}{\rightarrow} \mathcal{O}_{\mathbb{P}^5}(m) \rightarrow
\mathcal{O}_{\mathcal{X}}(m),
\]
and
\[
\mathcal{O}^{\oplus 4}_{\mathbb{P}^5}(-1)  \stackrel{\psi_{\pm}}{\rightarrow} \mathcal{O}^{\oplus 4}_{\mathbb{P}^5}
\rightarrow S_{\pm},
\] 
where $S_+$ and $S_-$ are the two spinor bundles\footnote{See \cite{ottaviani1988} for the basics of spinor bundles on quadrics.} of the quadric $\mathcal{X}$, we see the generator $\mathcal{A}_m$ becomes $\pi^*\mathcal{O}_{\mathcal{X}}(m)$, and $\mathcal{F}_{\pm}$ becomes $\pi^*S_\pm$ under the RG-flow at $r \gg 1$, where $\pi$ is the projection map of the canonical line bundle of $Gr(2,4) \cong \mathcal{X}$ onto the base. The spinor bundles on $Gr(2,4)$ are nothing but the tautological bundle $S$ and the (dual) universal quotient bundle $Q^{\vee}$ (see \cite{addington2011spinor}). Due to the isomorphism
\[
Q^{\vee} \cong \det Q^{\vee} \otimes Q \cong \det S \otimes (S\rightarrow \mathcal{O}^{\oplus 4}) ,
\]
we can exchange the generators $\pi^* S$ and $\pi^* Q^{\vee}$ by $\pi^* S$ and $\pi^*(\det S \otimes (S\rightarrow \mathcal{O}^{\oplus 4}))$, hence the window category of the dual abelian model is equivalent to
\begin{equation}\label{equidecomp}
\begin{aligned}
&D(\mathrm{Tot}(\mathcal{O}_{\mathcal{X}}(-4))) = \langle \pi^*S_-, \pi^*S_+, \pi^*\mathcal{O}_X, \pi^*\mathcal{O}_X(1), \pi^*\mathcal{O}_X(2), \pi^*\mathcal{O}_X(3) \rangle \\
&\cong \langle \pi^* S, \pi^* (S \otimes \det S), \mathcal{O}, \pi^*\det S, \pi^*\det S^{\otimes 2}, \pi^*\det S^{\otimes 3} \rangle = D(K_{Gr(2,4)}),
\end{aligned}
\end{equation}
in agreement with the window category we found in Sec. \ref{sec:evenresol}.

Moreover, we expect that the category of B-branes in the $\xi\ll -1$ phase corresponds to a noncommutative resolution \cite{EH} of $CGr(k,N)$. Such a noncommutative resolution is actually studied in \cite{Doyle:2021bsi} (using the techniques of \cite{vspenko2015non}), where only the case $\mathrm{gcd}(k,N)=1$ was considered.
From the argument above we can infer that the noncommutative resolution of $CGr(2,4)$ must have the following semiorthogonal decomposition:
\begin{equation}\label{semiequiv}
\langle \mathrm{MF}_{\mathbb{Z}_2}(G), \mathcal{A}_0, \mathcal{A}_1, \mathcal{A}_2, \mathcal{A}_3 \rangle \cong D(K_{Gr(2,4)}),
\end{equation}
where $\mathrm{MF}_{\mathbb{Z}_2}(G)$ denotes the category of B-branes of the LG model on $\mathbb{C}^6/\mathbb{Z}_2$ with superpotential $G$. The equivalence
\[
\mathrm{MF}_{\mathbb{Z}_2}(G) \cong \langle \mathcal{F}_+, \mathcal{F}_- \rangle
\]
can be seen from the following fact:
The category $\mathrm{MF}_{\mathbb{Z}_2}(G)$ appears as the category of B-branes of the Higgs branch in the LG phase of the $U(1)$ GLSM with six chiral fields $X_{ij}$ of charge 1 and one chiral field $P_1$ of charge $-2$ with superpotential $W = P_1 G(X)$. Note that this GLSM is anomalous and the LG phase has four Coulomb vacua apart from the Higgs branch.
The category $\mathrm{MF}_{\mathbb{Z}_2}(G)$ is therefore equivalent to the small-window category of this anomalous GLSM \cite{Clingempeel:2018iub}. With a suitable choice of the theta-angle, the small-window category of this model consists of B-branes whose $\rho_{M}$ has $U(1)$ gauge weights $q\in\{-1,0\}$. Therefore, it can be seen that the small-window category is generated by $\mathcal{F}_{\pm}$ given by the matrix factorizations \eqref{fplus} and \eqref{fminus}. According to Theorem 3.10 of \cite{Orlov:2005mnh}, the category $\mathrm{MF}_{\mathbb{Z}_2}(G)$ is equivalent to the triangulated category of singularities $\mathrm{D^{gr}_{Sg}}(A)$ of the graded $A$-modules, where $A = \mathbb{C}[X_{12},\cdots, X_{34}]/{\langle G(X) \rangle}$ is the coordinate ring of the affine cone $CGr(2,4)$. This is an evidence that the category of B-branes in the $\xi \ll -1$ phase with semiorthogonal decomposition \eqref{semiequiv} gives rise to a noncommutative resolution of the $\mathbb{Z}_4$-orbifold of $CGr(2,4)$.

\section*{Acknowledgement}

We thank C. Brav, W. Donovan, B. Lin, D. Pomerleano and E. Scheidegger
for helpful and enlightening discussions. MR thanks R. Eager, K. Hori and J. Knapp for collaboration on related projects. In particular we thank R. Eager, K. Hori for taking a detailed look and helping correct some mistakes in the first version of this draft. JG was supported by the National Natural Science Foundation of China (Grant No. 12475005), the Natural Science Foundation of Shanghai Municipality (Grant No. 24ZR1468600), and the Fundamental Research Funds for the Central Universities. MR acknowledges support from the
National Key Research and Development Program of China, Grant No.
2020YFA0713000. MR
also acknowledges Higher School of Economics, Simons Center for Geometry
and Physics, St. Petersburg State University and UC Berkeley for
hospitality at the final stages of
this work.


\appendix
\section{Generating empty branes from the Euler sequence}
\label{EmptGen}
The following example will illustrate how one arrives at the formula for generating empty branes from the Euler sequence on the Grassmannian, and how one finds the corresponding Young tableau $\lambda$ for a chosen $\mu$.

A useful example is that of $\mu=(2,1)$ for $K_{Gr(2,5)}$. The Schur functor on the Euler sequence is
\begin{align}
	L_{(2,1)} \left[ S \rightarrow V \rightarrow Q \right].
\end{align}
Given that the sequence in the brackets is exact, so too is
\begin{align}
	L_{(2,1)} \left[ S \rightarrow V \right] \rightarrow L_{(2,1)} \left[ Q
\right].
\end{align}
The second object involves those pesky sheaves constructed from $Q$, so the following steps are taken to convert them to sheaves constructed from $S^\vee$.

First, take a power of $Q$ equal to the total number of boxes in $\mu$. In this case,
\begin{align}
	Q \otimes Q \otimes Q \cong L_{(3)} Q \oplus 2 L_{(2,1)} Q \oplus L_{(1,1,1)} Q.
\end{align}
Due to
\begin{align}
	\wedge^k E \cong \wedge^r E \otimes \wedge^{r-k} E^\vee
	\label{WedgeIdent}
\end{align}
the power of $Q$ is also isomorphic to
\begin{align}
	Q^{\otimes 3} &\cong \left(\det Q \otimes \wedge^2 Q^\vee \right)^{\otimes 3} \\
	&\cong (\det Q )^{\otimes 3} \otimes \left(L_{(3,3)} Q^\vee \oplus 2L_{(3,2,1)}Q^\vee \oplus L_{(2,2,2)} Q^\vee \right).
\end{align}
By equating the coefficients and/or ranks of the bundles, one can infer the relation
\begin{align}
	L_{(2,1)}Q \cong (\det Q )^{\otimes 3} \otimes L_{(3,2,1)} Q^\vee.
	\label{RelEx}
\end{align}
The $\lambda$ here is then the Young tableau $(3,2,1)$ for $\mu=(2,1)$.

Occasionally, the relation can be simplified further. For this example, $L_{(3,2,1)} Q^\vee$ can be decomposed into
\begin{align}
	L_{(3,2,1)} Q^\vee &\cong L_{(1,1,1)} Q^\vee \otimes L_{(2,1)} Q^\vee \\
	&\cong \det Q^\vee \otimes L_{(2,1)} Q^\vee,
\end{align}
reducing Eq. \eqref{RelEx} to
\begin{align}
	L_{(2,1)}Q \cong (\det Q )^{\otimes 2} \otimes L_{(2,1)} Q^\vee.
\end{align}

The relations $\det Q \cong \det S^\vee$ and $Q \cong S \rightarrow V$ (from the Euler sequence) can then be utilised to replace sheaves of $Q$ for $S$:
\begin{align}
	(\det S^\vee )^{\otimes 2} \otimes L_{(2,1)^{T}} (V^\vee \rightarrow
S^\vee).
\end{align}
The Schur functor on the Euler sequence for this example then becomes
\begin{align}
	L_{(2,1)}(S \rightarrow V) \rightarrow (\det S^\vee)^{\otimes 2} \otimes
L_{(2,1)^{T}} (V^\vee \rightarrow S^\vee ).
\end{align}

\section{Examples of Generating Empty Branes for Monodromy Actions}
\label{Examples}

In this section we collect two examples, namely the cases $K_{G(2,5)}$ and $K_{G(2,6)}$, to illustrate how shifting between windows in the $\xi\gg 1$ phase can be performed, using only empty branes constructed from exact sequences of the form $L_{\mu}E$, where $E$ is the Euler sequence.

\subsection{$K_{G(2,5)}$}
\label{Examples25}

Consider an arbitrary B-brane $\mathcal{B}$ with charge configuration belonging to the window $\omega_{-2}$ in Fig. \ref{G25MonWins}. To undergo monodromy for a loop around $\theta=-\pi$ we must window shift through $\omega_{-1}$ (shown also in Fig. \ref{G25MonWins}). The charge needed to be replaced is on the $n=0$ shifted diagonal. Specifically, we need to replace the module $\mathcal{W}_{(3,3)}$. In a first step, we need to do it using an empty brane in the $\xi\ll-1$ phase, namely (as pointed out in Sec. \ref{OddNBraneGen}):
\begin{align}
	\mathcal{E}^{-}_{0} = \mathcal{W}_{(2,2)} \rightarrow \mathcal{W}_{(-3,-3)}.
\end{align}
Upon a twist by $\mathcal{W}_{(1,1)}$, it serves our purpose: taking multiple cones with $\mathcal{B}$ will restrict $\mathcal{B}$ to $\omega_{-1}$.

Finally we have to return to $\omega_{-2}$, but using empty branes in the $\xi\gg 1$ phase. So, we need to `replace back' $\mathcal{W}_{(-2,-2)}$ by branes with charges in $\omega_{-2}$. This can be done by taking cones with ($\mathcal{W}_{(1,1)}$ twists of) the UV lift of:
\begin{align}\label{emptybrane25}
	\pi^*L_{(2,2)} E,
\end{align}
since this empty brane will have a copy of $\mathcal{W}_{(2,2)}$ on one end. Indeed, it is equivalent to
\begin{align}\label{LEEE}
	\begin{array}{c@{}c@{}c@{}c@{}c@{}c@{}c@{}c@{}c@{}c@{}c@{}c@{}c@{}c@{}c@{}}
		\mathcal{W}_{(2,2)} & \rightarrow & \mathcal{W}_{(2,1)}^{\oplus 5} & \rightarrow & \mathcal{W}_{(1,1)}^{\oplus 15} \oplus \mathbfcal{W}_{(2,0)}^{\oplus 10} & \rightarrow & \mathcal{W}_{(1,0)}^{\oplus 40} & \rightarrow & \mathcal{W}_{(0,0)}^{\oplus 50} & \rightarrow & \mathcal{W}_{(-2,-2)}^{\oplus 15} & \rightarrow & \mathcal{W}_{(-2,-3)}^{\oplus 5} & \rightarrow & \mathcal{W}_{(-3,-3)}
	\end{array}.
\end{align}
However, the module in bold in (\ref{LEEE}), namely $\mathbfcal{W}_{(2,0)}$ does not fit the window $\omega_{-1}$ (also upon twisting by $\mathcal{W}_{(1,1)}$). Therefore we need to keep taking cones by other empty branes, in this case, a logical candidate will be the UV lift the sequence $\pi^*(\wedge^2 V \otimes L_{(1,1)} E)$, i.e.:
\begin{align}
	\left[ \mathbfcal{W}_{(2,0)} \rightarrow \mathcal{W}_{(1,0)}^{\oplus 5} \rightarrow \mathcal{W}_{(0,0)}^{\oplus 10} \rightarrow \mathcal{W}_{(-1,-1)}^{\oplus 5} \rightarrow \mathcal{W}_{(-1,-2)} \right]^{\oplus 10}.
\end{align}
Afterwards there are no more branes left with charges outside $\omega_{-2}$. This means, in order to shift between the window $\omega_{-1}$ to $\omega_{-2}$ in the $\xi\gg 1$ phase, we only need to take multiple cones by the UV lifts of $\pi^*(L_{(2,2)} E\otimes S^{(1,1)})$ and $\pi^*(L_{(1,1)} E)$, i.e. a finite number of empty branes.



Now  consider a B-brane $\mathcal{B}$ grade restricted to $\omega_{-2}$, but we undergo monodromy through $\theta=-2\pi$. Then we need to go through the window $\omega_{-3}$. The charges needed to be replaced are on the $n=1$ shifted diagonal and corresponds to the charges of $\mathcal{W}_{(-1,-2)}$.

This can be done in the $\xi\ll -1$ phase by using the empty brane:
\begin{align}
	\mathcal{E}^{-}_{1} = \mathcal{W}_{(2,1)} \rightarrow \mathcal{W}_{(-2,-3)},
\end{align}
upon twisting by $\mathcal{W}_{(1,1)}$. For the window shift, from $\omega_{-3}$ back to $\omega_{-2}$ in the positive phase, we proceed likewise. First we take cones by (the $\mathcal{W}_{(1,1)}$ twist of) the UV lift of
\begin{align}
	\pi^*L_{(2,1)} E,
\end{align}
which is equivalent to the sequence
\begin{equation}\label{OddNPiece1}
	\begin{tikzcd}
  \mathcal{W}_{(2,1)}  \rightarrow  \mathcal{W}_{(1,1)}^{\oplus 5} \oplus \mathbfcal{W}_{(2,0)}^{\oplus 5}  \rightarrow  \mathcal{W}_{(1,0)}^{\oplus 25}  \rightarrow  \mathcal{W}_{(0,0)}^{\oplus 40}   \ar[draw=none]{d}[name=X, anchor=center]{}
   \ar[rounded corners,
            to path={ -- ([xshift=2ex]\tikztostart.east)
                      |- (X.center) \tikztonodes
                      -| ([xshift=-2ex]\tikztotarget.west)
                      -- (\tikztotarget)}]{d}[at end]{} \\      
  \mathcal{W}_{(-2,-2)}^{\oplus 40} \rightarrow \mathcal{W}_{(-2,-3)}^{\oplus 25}  \rightarrow  \mathbfcal{W}_{(-2,-4)}^{\oplus 5} \oplus \mathcal{W}_{(-3,-3)}^{\oplus 5}  \rightarrow  \mathcal{W}_{(-3,-4)}.
\end{tikzcd}
\end{equation}
This time, we marked two modules in bold in (\ref{OddNPiece1}) that do not belong to $\omega_{-2}$. There are two more empty branes required to remove these. On one hand we have the exact sequence:
\begin{align}
	\pi^*\left(V \otimes L_{(1,1)} E \right),
\end{align}
which is equivalent to
\begin{align}
	\pi^*\left[ \mathbf{S^{(2)}} \rightarrow S^{\oplus 5} \rightarrow \mathcal{O}^{\oplus 10} \rightarrow S^{\vee(1,1) \oplus 5} \rightarrow S^{\vee(2,1)} \right]^{\oplus 5},
	\label{OddNPiece2}
\end{align}
and on the other hand we have the exact sequence
\begin{align}
	&\pi^*\left[V \otimes \mathrm{det}^2 S \otimes L_{(1,1)} E\right]^\vee \nonumber \\
=&\pi^*\left[ S^{(2,2)} \otimes \left( \mathbf{S^{(2)}} \rightarrow S^{\oplus 5} \rightarrow \mathcal{O}^{\oplus 10} \rightarrow S^{\vee(1,1) \oplus 5} \rightarrow S^{\vee(2,1)} \right) \right]^{\vee \oplus 5},
\end{align}
or, equivalently
\begin{align}
	 \pi^*\left[ S^{\vee} \rightarrow S^{\vee(1,1) \oplus 5} \rightarrow S^{\vee(2,2) \oplus 10} \rightarrow S^{\vee(3,2) \oplus 5} \rightarrow \mathbf{S^{\vee(4,2)}} \right]^{\oplus 5}.\label{OddNPiece3}
\end{align}
Afterwards there are no more bundles left, with charges outside $\omega_{-2}$. This means, in order to shift between the window $\omega_{-3}$ to $\omega_{-2}$ in the $\xi\gg 1$ phase, we only need to take multiple cones with the UV lifts of $\pi^*\left(L_{(2,1)} E\otimes S^{(1,1)}\right)$, $\pi^*\left(L_{(1,1)} E\otimes S^{(1,1)}\right)$ and\\ $\pi^* \left[ \left(\mathrm{det}^2 S \otimes L_{(1,1)} E\right)^{\vee}\otimes S^{(1,1)}\right]$.


\subsection{$K_{G(2,6)}$}\label{subsecG26}

In this section we perform the monodromy for $K_{G(2,6)}$. We consider a B-brane $\mathcal{B}$ grade-restricted to the window $\omega'_{-3}$, as described in (\ref{primewindoww}). This window is shown in the top left diagram of Fig. \ref{G26PossWins}. We will start by taking a loop around $\theta=-2\pi$ through th window $\omega_{-2}$. Note we can choose $\omega'_{-2}$ or any other valid window configuration we want. In any case, in order to shift from $\omega'_{-3}$ to $\omega_{-2}$ we need to exchange the module $\mathcal{W}_{(4,4)}$ by $\mathcal{W}_{(-2,-2)}$. Their weights belong to the $n=0$ shifted diagonal so, in order to map $\mathcal{B}$ to $\omega_{-2}$, using an empty brane in the $\xi\ll -1$ phase, we need to take successive cones with:
\begin{align}
	\mathcal{E}^{-}_{0}=\mathcal{W}_{(2,2)} \rightarrow \mathcal{W}_{(-4,-4)},
\end{align}
upon twisting it by $\mathcal{W}_{(2,2)}$. Finally we need to return to $\omega'_{-3}$ using an empty brane in the $\xi\gg 1$ phase. We can do this by taking cones with the UV lift of (upon twisting it by $\mathcal{W}_{(2,2)}$)
\begin{align}
	\pi^*L_{(2,2)} E.
\end{align}
This is equivalent to
\begin{equation}\label{EvenNn0Piece1}
	\begin{tikzcd}
  \mathcal{W}_{(2,2)}  \rightarrow  \mathcal{W}_{(2,1)}^{\oplus 6}  \rightarrow  \mathcal{W}_{(1,1)}^{\oplus 21} \oplus \mathbfcal{W}_{(2,0)}^{\oplus 15}  \rightarrow  \mathcal{W}_{(1,0)}^{\oplus 70}  \rightarrow  \mathcal{W}_{(0,0)}^{\oplus 105}
\ar[draw=none]{d}[name=X, anchor=center]{}
   \ar[rounded corners,
            to path={ -- ([xshift=2ex]\tikztostart.east)
                      |- (X.center) \tikztonodes
                      -| ([xshift=-2ex]\tikztotarget.west)
                      -- (\tikztotarget)}]{d}[at end]{} \\      
  \mathcal{W}_{(-2,-2)}^{\oplus 105}  \rightarrow \mathcal{W}_{(-2,-3)}^{\oplus 70}   \rightarrow  \mathcal{W}_{(-2,-4)}^{\oplus 15} \oplus \mathcal{W}_{(-3,-3)}^{\oplus 21} \rightarrow \mathcal{W}_{(-3,-4)}^{\oplus 6} \rightarrow \mathcal{W}_{(-4,-4)}.
\end{tikzcd}
\end{equation}
The module $\mathcal{W}_{(2,0)}$, in bold, (or more precisely $\mathcal{W}_{(2,0)}\otimes \mathcal{W}_{(2,2)}=\mathcal{W}_{(4,2)}$) do not belong to $\omega'_{-3}$ but still maps $\mathcal{B}$ to another valid window inside $\hat{\omega}_{-3}$. If we insist on return to the configuration $\omega'_{-3}$ we just can take cones with the empty brane from the exact sequence $L_{(1,1)} E$ (twisted by $\mathcal{W}_{(2,2)}$):
\begin{align}
	\mathbfcal{W}_{(2,0)} \rightarrow \mathcal{W}_{(1,0)}^{\oplus 6} \rightarrow \mathcal{W}_{(0,0)}^{\oplus 15} \rightarrow \mathcal{W}_{(-1,-1)}^{\oplus 15} \rightarrow \mathcal{W}_{(-1,-2)}^{\oplus 6} \rightarrow \mathcal{W}_{(-1,-3)}.
\end{align}
Therefore we conclude that, in the $\xi\gg 1$ phase, we only need to take cones with the UV lifts of $\pi^*\left(L_{(2,2)}E\otimes S^{(2,2)}\right)$ and, possibly $\pi^*\left(L_{(1,1)}E\otimes S^{(2,2)}\right)$, if we insist to return to $\omega'_{-3}$.

Now we consider the case of encircling the singularities at $\theta=-3\pi$. We again consider $\mathcal{B}\in\omega'_{-3}$. In order to map it to $\omega_{-4}$ (for any valid choice of $\omega_{-4}$) we need to exchange the module $\mathcal{W}_{(-1,-2)}$ by $\mathcal{W}_{(5,4)}$. They belong to the $n=1$ shifted diagonal. Then such a map, in the $\xi\ll -1$ pahse, will be implemented by successive cones by $\mathcal{E}^{-}_{1}\otimes \mathcal{W}_{(2,2)}$ where
\begin{align}
	\mathcal{E}^{-}_{1}=\mathcal{W}_{(2,1)} \rightarrow \mathcal{W}_{(-4,-5)}.
\end{align}

For the map $\omega_{-4}\rightarrow\omega'_{-3}$ in the $\xi\gg 1$ phase we can start by considering the UV lift of $\pi^*\left(L_{(2,1)} E\otimes S^{(2,2)}\right)$, where the UV lift of $\pi^*L_{(2,1)} E$ is explicitly given by
\begin{equation}\label{EvenNn1Piece1}
	\begin{tikzcd}
  \mathcal{W}_{(2,1)}  \rightarrow  \mathcal{W}_{(1,1)}^{\oplus 6} \oplus \mathbfcal{W}_{(2,0)}^{\oplus 6}  \rightarrow  \mathcal{W}_{(1,0)}^{\oplus 36}  \rightarrow  \mathcal{W}_{(0,0)}^{\oplus 70}  \rightarrow  \mathcal{W}_{(-2,-2)}^{\oplus 210} \rightarrow \mathcal{W}_{(-2,-3)}^{\oplus 210}   \ar[draw=none]{d}[name=X, anchor=center]{}
   \ar[rounded corners,
            to path={ -- ([xshift=2ex]\tikztostart.east)
                      |- (X.center) \tikztonodes
                      -| ([xshift=-2ex]\tikztotarget.west)
                      -- (\tikztotarget)}]{d}[at end]{} \\      
   \mathcal{W}_{(-3,-3)}^{\oplus 70} \oplus \mathcal{W}_{(-2,-4)}^{\oplus 90}  \rightarrow  \mathbfcal{W}_{(-2,-5)}^{\oplus 15} \oplus \mathcal{W}_{(-3,-4)}^{\oplus 36} \rightarrow  \mathcal{W}_{(-3,-5)}^{\oplus 6} \oplus \mathcal{W}_{(-4,-4)}^{\oplus 6}  \rightarrow  \mathcal{W}_{(-4,-5)}.
\end{tikzcd}
\end{equation}
The two modules in bold have different meaning. The module $\mathcal{W}_{(-2,-5)}$ does not belong to any valid subwindow of $\hat{\omega}_{-3}$, therefore it must be removed. We can do it by taking cones with the UV twist of (twisted by $\mathcal{W}_{(2,2)}$)
\begin{align}
\pi^*[S^{(2,2)}\otimes L_{(1,1,1)}E]^{\vee},
\end{align}
where
\begin{align}
		L_{(1,1,1)} E = \mathbf{S^{(3)}} \rightarrow \underline{S^{(2) \oplus 6}} \rightarrow S^{\oplus 15} \rightarrow \mathcal{O}^{\oplus 20} \rightarrow S^{\vee(1,1) \oplus 6} \rightarrow S^{\vee(2,1)}.
\end{align}
To other module in bold in (\ref{EvenNn1Piece1}), i.e. the module $\mathcal{W}_{(2,0)}$ needs only to be removed if we insist on keeping the original shape of the window $\omega'_{-3}$. As in the case of monodromy around $\theta=-2\pi$, we can just use the empty brane from the exact sequence $L_{(1,1)} E\otimes S^{(2,2)}$ for this purpose. In summary, monodromy around $\theta=-3\pi$ requires us to take cones, in the $\xi\gg 1$ phase, with the UV lifts of $\pi^*\left(L_{(2,1)}E\otimes S^{(2,2)}\right)$, $\pi^*\left[(\mathrm{det}^{2}S\otimes L_{(1,1,1)}E)^{\vee}\otimes S^{(2,2)}\right]$ and, possibly $\pi^*\left(L_{(1,1)}E\otimes S^{(2,2)}\right)$, if we insist to return to $\omega'_{-3}$.

\bibliographystyle{fullsort}
\bibliography{kgref.bib}
\end{document}